\documentclass[a4paper,11pt]{article}
\pdfoutput=1 % if your are submitting a pdflatex (i.e. if you have
\usepackage{jcappub} % for details on the use of the package, please see the JINST-author-manual
\usepackage{aas_macros}
\usepackage{subcaption}
\usepackage[normalem]{ulem}
\usepackage[dvipsnames]{xcolor}
\usepackage{tablefootnote}
\usepackage{orcidlink}

\newcommand{\codename}{\textsc{Disco-Dj}}
\newcommand{\codenameEB}{\textsc{Disco-EB}}

\newcommand{\dd}{\ensuremath{\text{d}}}

\arxivnumber{2311.03291} % Only if you have one
\title{\boldmath DISCO-DJ I: a differentiable Einstein-Boltzmann solver for cosmology}
\author[a,b]{Oliver Hahn\orcidlink{0000-0001-9440-1152}}
\author[a]{Florian List\orcidlink{0000-0002-3741-179X}}
\author[c]{Natalia Porqueres\orcidlink{0000-0002-7599-966X}}

\affiliation[a]{Department of Astrophysics, University of Vienna,\\ Türkenschanzstraße 17, 1180 Vienna, Austria}
\affiliation[b]{Department of Mathematics, University of Vienna,\\ Oskar-Morgenstern-Platz 1, 1090 Vienna, Austria}
\affiliation[c]{Department of Physics, University of Oxford,\\ Denys Wilkinson Building, Keble Road, Oxford OX1 3RH, UK}
\emailAdd{oliver.hahn@univie.ac.at}

\abstract{We present the Einstein-Boltzmann module of the \codename{} (\textbf{DI}fferentiable \textbf{S}imulations for \textbf{CO}smology -- \textbf{D}one with \textbf{J}\textsc{ax}) software package. This module implements a fully differentiable solver for the linearised cosmological Einstein-Boltzmann equations in the \textsc{Jax} framework, and allows computing Jacobian matrices of all solver output with respect to all input parameters using automatic differentiation. This implies that along with the solution for a given set of parameters, the tangent hyperplane in parameter space is known as well, which is a key ingredient for cosmological inference and forecasting problems as well as for many other applications. We discuss our implementation and demonstrate that our solver agrees at the per-mille level with the existing non-differentiable solvers \textsc{Camb} and \textsc{Class}, including massive neutrinos and a dark energy fluid with parameterised equation of state. We illustrate the dependence of various summary statistics in large-scale structure cosmology on model parameters using the differentiable solver, and finally demonstrate how it can be easily used for Fisher forecasting, with a forecast for Euclid as an example. Since the implementation is significantly shorter and more modular than existing solvers, we believe it will be more straightforward to extend our solver to include additional physics, such as additional dark energy and dark matter models, modified gravity, or other non-standard physics in the future.}

\begin{document}
\maketitle
\flushbottom

%%%%%%%%%%%%%%%%%%%%%%%%%%%%%%%%%%%%%%%%%%%%%%%%%%%%%%%%%%%%%%%%%
\section{Introduction}
\label{sec:intro}
Describing the evolution of perturbations of Einstein's field equations in a cosmological setting is an almost century-old quest and the foundation of inhomogeneous cosmology as much as that of all of extragalactic astrophysics. These perturbations, leading to deviations from the homogeneous Friedmann-Lema\^itre-Robertson-Walker (FLRW) spacetimes and the growth of structure, are described by the Einstein-Boltzmann equations, which are a set of coupled partial differential equations that describe the evolution of the energy density of the various components of the Universe along with the perturbations to the FLRW metric. First derivations of the equations date back many decades \cite{Lifshitz:1946,LifshitzKhalatnikov:1963} and subsequent revisions provide the foundation of all of modern inhomogeneous cosmology \cite{Weinberg:1972gravitation,Peebles:1980LSS,MaBertschinger:1995}.   

The Einstein-Boltzmann equations are a cornerstone of modern cosmology, and are the basis for the computation of the cosmic microwave background (CMB) anisotropies, the matter power spectrum, and other observables. The matter power spectrum, or its constituents, the individual transfer functions of the various matter components, is the starting point for all predictions of large-scale structure observables, such as the galaxy n-point spectra, weak lensing spectra, the halo mass function (in analytic models), and in particular also the initial conditions for all cosmological simulations. Building on the {\em linear} Einstein-Boltzmann equations, the {\em non-linear} evolution of structure is then typically described in the non-relativistic collisionless limit by the Vlasov-Poisson system of equations, which is the basis for all non-linear perturbation theory and all $N$-body simulations (e.g. \cite{Peebles:1980LSS,Bernardeau:2002review,AnguloHahn:2022review}).

Several famous software packages have been developed that integrate the linearised Einstein-Boltzmann equations. The most commonly used (historically) include the \textsc{Cosmics/Linger} code \cite{MaBertschinger:1995}, followed by a version specifically optimised for CMB predictions, the \textsc{CMBfast} code \cite{CMBfast}, which included the  line-of-sight approximation which allowed a dramatically faster computation of CMB spectra. The increasing requirements of the WMAP \cite{WMAPmission:2003} and Planck \cite{Planckmission:2011} CMB missions for efficient codes have led to the highly optimised and accurate \textsc{CAMB}\footnote{\url{https://camb.info}} \cite{CAMB} and \textsc{CLASS}\footnote{\url{https://lesgourg.github.io/class_public/class.html}} \cite{CLASS,CLASS_approx} codes. A recent addition is \textsc{PyCosmo}\footnote{\url{https://cosmology.ethz.ch/research/software-lab/PyCosmo.html}} \cite{Refregier:2018} which relies on automatic translation of symbolic expressions to compiled code. Both \textsc{Camb} and \textsc{Class} feature a large number of optimizations (e.g. approximate asymptotic forms of the equations valid in various regimes, robust high-order implicit integrators), are highly optimised, and are written in compiled functional languages -- \textsc{Fortran} (CAMB) and \textsc{C} (CLASS) -- but both now have convenient \textsc{Python} interfaces.

Both packages share a common minimal set of implemented equations (see \cite {MaBertschinger:1995, Bucher:2000}), which model the evolution of cold dark matter (CDM), baryons, photons, massless and massive neutrinos, a dark energy fluid, as well as metric perturbations. While already the first codes included these energy components, \textsc{Camb} and \textsc{Class} have by now been extended to include a plethora of additional models, including various dark energy and modified gravity models \cite{EFTgravityCAMB:2014,HiClass:2017,MGCAMB:2019}, and other variations such as scalar field dark matter \cite{AxionCAMB:2015,Foidl:2022}, early dark energy \cite{Hill:2020}, and many more.

In this article, we present the Einstein-Boltzmann (EB) module of a new software package, \codename{} (\textbf{DI}fferentiable \textbf{S}imulations for \textbf{CO}smology -- \textbf{D}one with \textbf{J}\textsc{ax}), that implements a fully differentiable Einstein-Boltzmann solver in the \textsc{Jax} framework \cite{JAX}. This solver is based on the \textsc{Diffrax} \textsc{Python} package \cite{Kidger:2021}, which (among others) implements a high-order implicit Runge-Kutta solver for stiff ordinary differential equations (ODEs). All code (including the \textsc{Diffrax} package) is built on top of the \textsc{Jax} framework, which provides automatic differentiation of \textsc{Python} code, and thus allows computing the Jacobian matrix of the Einstein-Boltzmann system as well as functions of its results with respect to all input parameters. This is a key ingredient for many applications relying on gradients, including cosmological inference and the computation of the Fisher matrix computation for forecasting or optimal data compression.

In addition to differentiability, the \textsc{Jax} framework is also optimised for execution on graphics processing units (GPUs), which allows for a significant speed-up of the computation of the Einstein-Boltzmann system, in particular when batching over multiple cosmological models, which is an embarrassingly parallel calculation.

The current version of the EB module of \codename{} (referred to as \textsc{Disco-Eb} in what follows) has a very lean and modular code base and is thus easy to extend and modify beyond its current minimal implementation from a programming point of view. In particular, we hope that in the future additional physics modules, such as dark energy models, modified gravity, or other non-standard physics will be added by the community. Eventually, we expect that also other deliverables beyond what is shown in this paper will be added, such as e.g. the CMB anisotropies or the lensing potential. Further, the EB module will be connected to a fast $N$-body solver, which we are also developing as part of the \codename{} package and which allows the inclusion of non-linear effects in the computation of the matter power spectrum and other key statistics and predictions for large-scale structure cosmology. We will present this in a future publication.

The structure of this article is as follows: In \autoref{sec:implementation} we discuss the implementation of the Einstein-Boltzmann solver in \codename, giving attention to differentiable solvers, implicit integrators for stiff ODEs, approximation schemes, and the thermal history solver. In \autoref{sec:validation} we validate the accuracy of the solver against the commonly used \textsc{Camb} and \textsc{Class} codes. We then discuss various applications of a differentiable Einstein-Boltzmann solver, including the computation of the dependence of various summary statistics on cosmological parameters, and the computation of the Fisher information matrix in \autoref{sec:applications}. Finally, we summarise and conclude in \autoref{sec:conclusions}. An overview of the equations implemented in the Einstein-Boltzmann module of \codename{} is given in \autoref{sec:governing_equations}.

%%%%%%%%%%%%%%%%%%%%%%%%%%%%%%%%%%%%%%%%%%%%%%%%%%%%%%%%%%%%%%%%%
\section{Implementation}
\label{sec:implementation}

In this section, we discuss aspects of the implementation of the linear Einstein-Boltzmann module of \codename.
%%%%%%%%%%%%%%%%%%%%%%%%%%%%%%%%%%%%%%%%%%%%%%%%%%%%%%%%%%%%%%%%%
\subsection{Overview}
The Einstein-Boltzmann equations are a set of non-linear coupled partial differential equations that describe the evolution of the energy density of the various components of the Universe along with the perturbations to the FLRW metric. A fully self-consistent non-linear solution all the way to late times and small scales is however so far computationally untractable (but see \cite{Pettinari:2013SONG} for a second-order EB solver). Instead, in cosmology, one typically splits the evolution into a linear and a non-linear part. The linear part is described by the linearised Einstein-Boltzmann equations. The non-linear part is typically described in the non-relativistic collisionless limit by the Vlasov-Poisson system of equations, which is solved with $N$-body methods (e.g. \cite{AnguloHahn:2022review}),  using input from the linear Einstein-Boltzmann solver.

In this work, we focus on the linear Einstein-Boltzmann equations including CDM, baryons, photons, massive and massless neutrinos, and clustering dark energy. The governing equations are stated in full in \autoref{sec:governing_equations} for reference. The same equations are also implemented in the popular \textsc{Camb} and \textsc{Class} solvers. Those solvers also include additional physical models (e.g. early dark energy, dark radiation, etc.) that we do not include (yet).

The new aspect of our work is that we implement the Einstein-Boltzmann equations in the \textsc{Jax} framework, which allows us to compute the Jacobian matrix of the Einstein-Boltzmann system (which is crucial input for implicit ODE solvers) as well as of any function of its results with respect to all parameters. For this reason, in this article, we first discuss the aspect of automatic differentiation in \autoref{sec:autodiff} below. The linearised Einstein-Boltzmann equations are so-called `stiff' coupled ODEs, which are most efficiently solved by implicit integrators, an aspect we discuss in \autoref{sec:stiff_odes}. Solvers such as \textsc{Camb} and \textsc{Class} employ a number of numerical approximation schemes in order to reduce the large number of equations needed to be solved and optimise the time-to-result for a given precision, which is critical for the use in data analysis. We discuss these aspects in \autoref{sec:approx}. Finally, we also discuss our choice of thermal history solver and its performance in \autoref{sec:thermal_history}.

%%%%%%%%%%%%%%%%%%%%%%%%%%%%%%%%%%%%%%%%%%%%%%%%%%%%%%%%%%%%%%%%%
\subsection{Automatic differentiation}
\label{sec:autodiff}
The autodiff paradigm \cite{Wengert:1964} has seen rapidly increasing interest recently due to the machine learning revolution. Traditionally, the computer-aided computation of derivatives could be subdivided into \emph{symbolic} and \emph{numerical} differentiation methods. Symbolic differentiation involves algorithmically determining the derivative of a function as a new symbolic expression by applying differentiation rules such as the chain rule, the product rule, etc. This approach is implemented in software packages such as the popular \textsc{Mathematica} program \cite{Mathematica} and the \textsc{Python} module \textsc{SymPy} \cite{Sympy}. While these programs are often able to find surprisingly simple expressions for derivatives that would be hard to compute by hand, more complex functions may result in lengthy expressions that are far from computationally optimal when being evaluated with concrete numerical values.
On the other hand, numerical differentiation relies on finite difference approximations of derivatives. This introduces a hyperparameter, namely the step size, upon which the results can depend sensitively. In particular, numerical approximations of higher derivatives are oftentimes simply too noisy to be useful in practice.

More recently, the autodiff paradigm (also known as \emph{automatic} differentiation) has become a popular third contender among computational differentiation methods and is now the de-facto standard in the field of machine learning. Automatic differentiation exploits the fact that programs perform the evaluation of any mathematical function of some inputs as a nested sequence of simple functions such as summation, multiplication, exponentiation, etc. The derivative of each of these building blocks is known, so when the program sequentially propagates the inputs through these blocks, the derivative information can be passed on alongside the computed values simply by using the chain rule. This particular flavour of autodiff is known as forward-mode differentiation. Alternatively, if the chain rule is instead performed starting at the outputs of the function and propagating backwards to the inputs, one obtains reverse-mode differentiation. Which of the two is computationally cheaper mainly depends on the dimensionality of the inputs and outputs (backward-mode for many inputs $\mapsto$ few outputs, e.g. when computing a scalar loss function w.r.t. the millions of parameters of a neural network, forward-mode for few inputs $\mapsto$ many outputs, e.g. when differentiating a field quantity w.r.t. a cosmological parameter). Keeping track of the derivatives with autodiff only entails a constant computational overhead for each of the function's building blocks. 

In our case, the quantity of interest, namely the matter power spectrum $P$, is obtained as the solution of a system of differential equations, specifically the cosmological Einstein-Boltzmann equations. Thus, the computation of gradients such as $\partial P / \partial \Omega_m$ requires us to differentiate through the solution of the differential equations. This can be done in two different ways: (1) differentiate iteratively through the time steps performed by the numerical solver (``discretise-then-optimise'' approach) or (2) set up a continuous ``adjoint equation'' for $\partial P / \partial \Omega_m$ that then must be solved backwards in time, also with a numerical method (``optimise-then-discretise'' approach).\footnote{Instead of the power spectrum itself, one could similarly differentiate a \emph{loss function} that gauges the match between the computed power spectrum $P$ and an observation $P_{\text{obs}}$ w.r.t. the cosmological parameters, which is a typical scenario in the context of parameter inference.} Generally, the optimise-then-discretise approach yields less accurate derivatives because the numerical discretisation of the backwards-in-time equations required for the computation of the derivatives is not synchronised with that for the forward-in-time equations for the quantities themselves \cite{onken2020discretize}; however, it is more memory-efficient than its discretise-then-optimise counterpart. 

In this work, we rely on Google's array-based autodiff \textsc{Python} library \textsc{Jax} \cite{JAX}. \textsc{Jax} conveniently allows writing differentiable programs in high-level \textsc{Python} that are then compiled just-in-time (JIT) to execute efficiently on graphics or tensor processing units (GPUs, TPUs), making use of the fast \textsc{Xla} (Accelerated Linear Algebra) compiler. Syntactically, much of \textsc{Jax} follows the syntax of \textsc{NumPy}, making it easy for users with \textsc{Python} experience to switch from plain \textsc{NumPy} (which runs on central processing units, i.e. CPUs, and is neither compiled nor differentiable) to \textsc{Jax}. It supports forward- and reverse-mode differentiation, the computation of higher-order derivatives such as Hessian matrices, holomorphic differentiation, and the vectorised computation of derivatives, among other useful features. The \textsc{Jax} ecosystem is steadily expanding, and there exist \textsc{Jax}-based packages for machine learning (e.g. \cite{flax2020github}), for physics simulations (e.g. \cite{kaymak2023end}), probabilistic programming (e.g. \cite{phan2019composable, bingham2019pyro}), and many other applications, which can be integrated seamlessly with custom \textsc{Jax} code. Importantly, we use the \textsc{Jax}-based \textsc{Diffrax} library for solving differential equations \cite{Kidger:2021}, which supports different flavours of ``discretise-then-optimise'' and ``optimise-then-discretise'' derivatives. The current version of \codenameEB{} supports both forward (via \textsc{Diffrax}'s \texttt{DirectAdjoint}) and backward (via \textsc{Diffrax}'s \texttt{BacksolveAdjoint}) differentiation. Other differentiable frameworks exist; notably e.g. \textsc{Bolt}\footnote{\url{https://github.com/xzackli/Bolt.jl}} implements a forward-differentiable Einstein-Boltzmann solver in \textsc{Julia} (but is unpublished at the time of writing).

In the past few years, \textsc{Jax} has garnered considerable interest in the cosmology community. For example, Ref.~\cite{campagne2023jax} presented a differentiable cosmology library using \textsc{Jax}, Ref.~\cite{Piras23} introduced a \textsc{Jax}-based inference framework for cosmology, and the $N$-body code by Ref.~\cite{pmwd} is fully differentiable thanks to its implementation in \textsc{Jax}. It has also been used in the context of estimating weak lensing shear \cite{Li:2023ehn} and for modelling the evolution of dark matter haloes \cite{Stevanovich:2023}.

%%%%%%%%%%%%%%%%%%%%%%%%%%%%%%%%%%%%%%%%%%%%%%%%%%%%%%%%%%%%%%%%%
\subsection{Stiff ODEs and implicit solvers}
\label{sec:stiff_odes}
The Einstein-Boltzmann equations, cf. \autoref{sec:governing_equations}, are a stiff system of ODEs due to the multiple time scales involved in the various physical processes, which makes them ill-suited for explicit time integration schemes \cite{hairer2010solving}. Various approximations can be employed (cf. \cite{CLASS_approx}, and \autoref{sec:approx}) to reduce the stiffness of the system (see also \cite{Nadkarni-Ghosh:2017}), but these approximations are generally not valid for the entire domain of integration. State-of-the-art solvers therefore use a combined approach of approximation schemes and a powerful high-order implicit time integration scheme \cite{CLASS_approx,CLASS_ncdm}. A main draw-back of implicit solvers is that the Jacobian matrix of the system needs to be computed and inverted at each time step, which is computationally expensive, particularly so for large systems such as the cosmological Einstein-Boltzmann equations. 

A main advantage of the autodiff paradigm is that the Jacobian matrix of the system can be computed automatically without specifying it explicitly or having to recourse to finite-difference approximations. For \codenameEB, we adopt the \textsc{Diffrax} package \cite{Kidger:2021} for this purpose, which comes with various high-order implicit and explicit integration and is also built on top of the \textsc{Jax} framework. Specifically, we adopt the highest order implicit solver implemented in \textsc{Diffrax}: the 7-stage {\sc Kvaerno5} solver \cite{kvaerno2004singly}, a singly diagonally implicit Runge-Kutta (ESDIRK) method of order 5/4. 

Due to the large number of equations, it is in principle critical to exploit the sparse structure of the Jacobian matrix for efficient implicit solvers -- such as done by \textsc{Class} which uses sparse storage for the Jacobian. However, sparse matrix operations in \textsc{Jax} are still in an experimental stage and thus, \textsc{Diffrax} does not (yet) support sparse matrices, which is the current performance bottleneck of \codenameEB{}.  

After linearisation of the Einstein-Boltzmann system \cite{MaBertschinger:1995}, the equations for individual wave numbers $k$ decouple in Fourier space and can thus be solved individually. The resulting Jacobian for multiple wave numbers thus has a block-diagonal structure. Since such sparse Jacobian matrices can currently not be handled, an optimised joint integration of multiple wave numbers, which leads to a block diagonal Jacobian (since the different modes do not couple), is currently not possible. Storing the full Jacobian for multiple modes is computationally prohibitive. While we are, therefore, currently restricted to integrating mode-by-mode in an embarrassingly parallel way, there is plausibly room for future performance improvements along such lines. In applications where the Einstein-Boltzmann system needs to be solved for multiple sets of cosmological parameters, e.g. in the context of inference tasks, it is straightforward to batch the computation over those sets in order to increase the computational speed, harnessing the highly parallel architecture of GPUs. Currently, we use however the \textsc{Jax} \texttt{vmap} function to vectorise the computation of multiple modes on the GPU.

%%%%%%%%%%%%%%%%%%%%%%%%%%%%%%%%%%%%%%%%%%%%%%%%%%%%%%%%%%%%%%%%%
\subsection{Approximation methods}
\label{sec:approx}
In order to speed up the integration of the Einstein-Boltzmann system, for \textsc{Class} and \textsc{Camb}, it has been critical to employ various approximation schemes \cite{CLASS_approx,CLASS_ncdm} in order to reduce the number of degrees of freedom of the high-dimensional system and to reduce the stiffness of the system. The situation is somewhat different for GPU-\textsc{Jax}-based solvers. While the Jacobian matrix of the system can be computed automatically (which is, in fact, more efficient than a finite difference calculation), the computation of the Jacobian matrix is still computationally expensive, and it is still desirable to reduce the number of degrees of freedom of the system. However, adaptively switching to a reduced set of variables based on a predefined criterion introduces additional algorithmic complexity, which is not desirable in the context of GPU-based solvers. Specifically, we found that neither of the commonly used approximation schemes yielded a $\mathcal{O}(1)$ performance improvement, and, while we implemented the tight coupling approximation \cite{MaBertschinger:1995,CLASS_approx} (valid when the baryons are fully ionised and tightly coupled to the photons), and all the approximations used in \textsc{Class} \cite{CLASS_approx} (notably the ultra-relativistic fluid approximation, applying to massless neutrinos at late times, as well as the radiation streaming approximation, applying to massless neutrinos and photons at even later times) in \codenameEB, we do not currently use any of them.

A main performance bottleneck lies in the computation of the Jacobian matrix, which scales quadratically with the number of equations (see \autoref{sec:governing_equations}). The number of equations scales with the number $\ell_\text{max}$ of the multipole where the photon and neutrino distribution functions are truncated, and the number of massive neutrino mass bins. Specifically we have
\begin{align}
    8 + 2 \times \ell_{\gamma,\text{max}} + \ell_{\nu \text{UR},\text{max}} + n_q \times \ell_{\nu \text{massive},\text{max}} 
\end{align}
equations. The `8' entails the equations for metric+CDM+baryon+dark energy perturbations, $\ell_{\gamma,\text{max}}$ is the maximum multipole moment of the photon distribution function, $\ell_{\nu \text{UR},\text{max}}$ is the maximum multipole moment of the massless neutrino distribution function, $n_q$ is the number of neutrino momentum bins, and $\ell_{\nu \text{massive},\text{max}}$ is the maximum multipole moment of the massive neutrino distribution function.

A key optimisation, therefore, lies in a clever choice of neutrino momentum bins so that $n_q$ can be chosen as small as possible. Specifically, \cite{Howlett:2012} found that a 3-5 bin Gauss-Laguerre inspired integral approximation used by \textsc{Camb} (see Appendix~A of \cite{Howlett:2012}) works very well in this context.  The neutrino momentum bins are chosen so that the following integral is well approximated by a sum over the bins
\begin{align}
    \frac{1}{4}\int_0^\infty {\rm d}q \frac{q^4 {\rm e}^q}{(1+{\rm e}^q)^2}\,v^w \,\nu_\ell \approx \sum_i K_i v^w \,\nu_\ell,
\end{align}
where $q$ is the neutrino momentum, $v$ is the (relativistic) neutrino velocity,  $\nu_\ell$ is a rescaled function of the $\ell$-th multipole moment of the neutrino distribution function, $w=-1,0,1$ (depending on which moment of the distribution function is to be computed, cf. eqs.~(55) of \cite{MaBertschinger:1995}), and $K_i$ are the kernel weights approximating the integral. The authors of \cite{Howlett:2012} propose the following very accurate binning schemes:
\begin{align*}
    \text{3-point:}&& q &= (0.913201,3.37517,7.79184),\\ & & K &= (0.0687359,3.31435,2.29911),\\
    \text{4-point:}&& q &= (0.7, 2.62814, 5.90428, 12),\\ & & K &= (0.0200251, 1.84539, 3.52736, 0.289427)\\
    \text{5-point:}&& q &= (0.583165, 2.0, 4.0, 7.26582, 13.0),\\ & & K &= (0.0081201, 0.689407, 2.8063, 2.05156, 0.12681)
\end{align*}
The truncation of the neutrino Boltzmann hierarchy $\ell_{\nu,\text{max}}= $ \texttt{lnumax} then determines the number of equations that need to be solved as the product of \texttt{lnumax} and the number of neutrino momentum bins. When neutrinos are sufficiently non-relativistic, \texttt{lnumax} can, in principle, be reduced to a very small number (2 or 3, cf. \cite{Howlett:2012}).

For our first proof-of-concept version of \codenameEB, we decided to leave out all such further approximations, but we expect that a possible factor of 2-3 in speed could be gained by a careful evaluation of the various approximation schemes on integration of the ODEs on GPUs (specifically optimizing how to best change the number of equations integrating directly in the integrator). We leave this for future work.

\subsection{Thermal history/recombination solver}
\label{sec:thermal_history}
\begin{figure}
    \centering
    \includegraphics[width=0.8\textwidth]{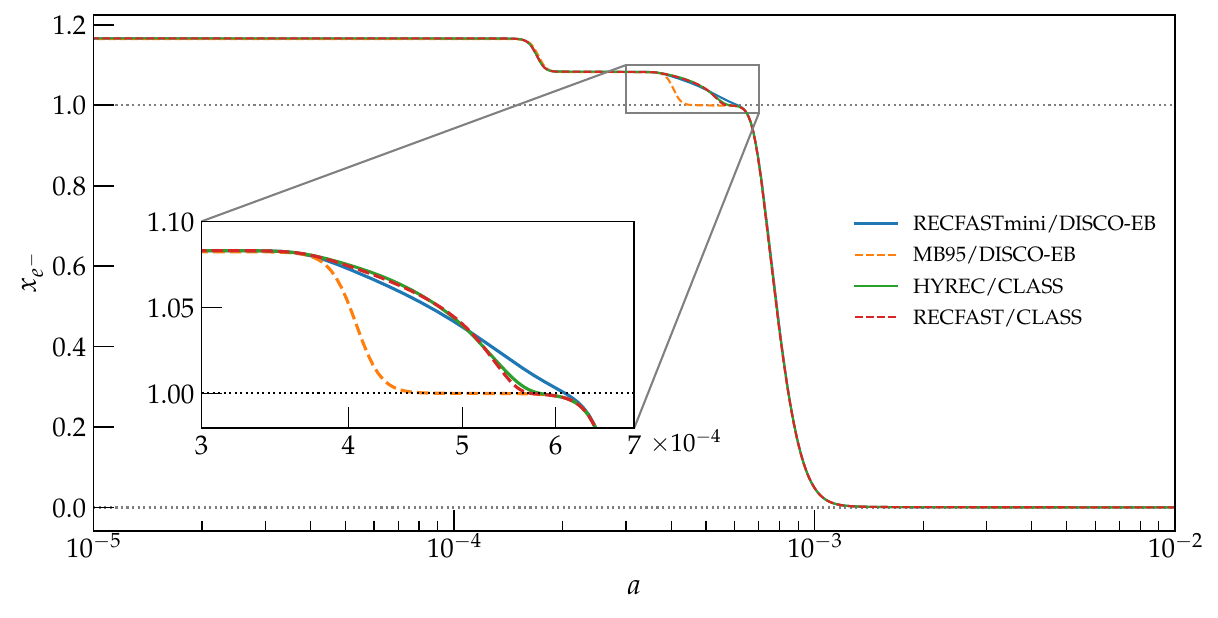}
    \caption{Performance of the simplified recombination solver `RECFASTmini' in \codename{} in comparison with the more accurate \textsc{RecFast} and \textsc{HyRec} solvers used in \textsc{Class}, and the original MB95 model \cite{MaBertschinger:1995} (also available in \codename). We show the ionization fraction $x_{e^-}$ as a function of the scale factor $a$ for the baseline cosmology computed using the simplified model based on the original \textsc{RecFast} (blue), the much older \textsc{MB95} model (orange dashed), as well the most recent \textsc{RecFast} (red dashed) and \textsc{HyRec} (green) modules in \textsc{Class}. The main difference is due to details in the treatment of the combined hydrogen-helium recombination. None of these differences (except MB95) have a measurable effect late-time on matter components.}
    \label{fig:thermal_history}
\end{figure}

Currently, we adopt a simplified thermal history solver in \codename, that is based on the original \textsc{RecFast} model \cite{Seager:1999recfast}. The original implementation of the thermal history in \textsc{Cosmics} (Section~5.8 of \cite{MaBertschinger:1995}, which is based on \cite{Peebles:1968recomb} with addition Helium) is not accurate enough to obtain sub-per-cent-level agreement with \textsc{Camb} or \textsc{Class}. All state-of-the-art Einstein-Boltzmann solvers employ at least \textsc{RecFast} \cite{Seager:1999recfast}, or the newer \textsc{CosmoRec} \cite{Chluba:2011cosmorec} and \textsc{HyRec} \cite{Ali-Haimoud:2011hyrec} (which has been further developed into \textsc{HyRec2} \cite{Lee:2020hyrec2}) solvers. We found however that, since we are predominantly interested in the matter power spectrum and late times (rather than a CMB spectrum), adopting the original \textsc{RecFast} model is accurate enough for our purposes. While the thermal history solver is thus not the main source of error in the late time matter power spectrum, we are planning to eventually add a more accurate recombination solver in \codename. The current version implements the original \textsc{RecFast} model, that we dub `RECFASTmini', as well as the much older \textsc{MB95} model \cite{MaBertschinger:1995}.

In \autoref{fig:thermal_history}, we compare the ionization history of the baseline cosmology computed using the simplified models in \codenameEB{} with the \textsc{RecFast} and \textsc{HyRec} solvers interfaced in \textsc{Class}. The main difference is due to the different treatment of the helium recombination, with a longer phase of He$^{+}$ recombination in \textsc{RecFast}. As we will show below, for late times $z\lesssim 1000$, the simplified `RECFASTmini' model is sufficient to reach agreement of late time spectra with \textsc{Class} or \textsc{Camb} to below one per cent.

%%%%%%%%%%%%%%%%%%%%%%%%%%%%%%%%%%%%%%%%%%%%%%%%%%%%%%%%%%%%%%%%%
\section{Validation and performance}
\label{sec:validation}
We next validate the performance of our solver against the \textsc{Camb} and \textsc{Class} solvers. First, we compare the precision of our solver for a given set of cosmological input parameters. Second, we demonstrate that our solver is able to compute the Jacobian matrix of the system with respect to all input parameters, and validate the correctness of the Jacobian matrix by comparing it to finite-difference approximations.

%%%%%%%%%%%%%%%%%%%%%%%%%%%%%%%%%%%%%%%%%%%%%%%%%%%%%%%%%%%%%%%%%
\subsection{Choice of fiducial cosmological model}
For all comparisons, unless otherwise indicated, we adopt a flat (extended) $\Lambda$CDM cosmology, with the cosmological parameter set shown in \autoref{tab:cosmo_params}, which is consistent with the \textsc{Planck 2018} \texttt{EE+BAO+SN} constraints \cite{Planck2018_cosmoparam}. Specifically, we consider the standard extension with inclusion of one massive neutrino, along with a dark energy fluid with parameterised (Chevallier-Polarski-Linder) equation of state $w = p/\rho = w_0 + w_a (1-a)$. We adopt $w_0=-0.99$ instead of $-1$ since derivatives w.r.t. $w_0$ are defined only in the one-sided limit at $w=-1$, a complication we thus avoid (although defining one-sided derivatives is, in principle, also supported by \textsc{Jax}).

%%%%%%%%%%%%%%%%%%%%%%%%%%%%%%%%%%%%%%%%%%%%%%%%%%%%%%%%%%%%%%%%%
\begin{table}[h!]
    \centering
    \begin{tabular}{l|l|l}
        \hline
        Parameter & Value & Meaning\\
        \hline
        $\Omega_\text{k}$ & $0.0$ & curvature density parameter (constraint)\\
        $\Omega_\text{b}$ & $0.0488911$ & baryon density parameter\\
        $\Omega_\text{m}$ & $0.3099$ & matter density parameter\\
        $h$ & $0.67742$ & Hubble parameter\\
        $n_\text{s}$ & $0.96822$ & primordial spectral index\\
        $A_\text{s}$ & $2.1064\times 10^{-9}$ & amplitude at $k_p=0.05\,\text{Mpc}^{-1}$\\
        $N_\text{eff}$ & $3.046$ & effective number of ultrarelativistic species\\
        $m_\nu$ & $0.06\,\text{eV}$ & mass of neutrino\\
        $w_0$ & $-0.99$ & dark energy equation of state parameter\tablefootnote{Note that we modified this from $-1$ as solutions are only one-sidedly differentiable at $w_0=-1$.}\\
        $w_a$ & $0.0$ & dark energy equation of state parameter\\
        $c_a^2$ & $1.0$ & dark energy sound speed\\
        $Y_\text{He}$ & $0.248$ & primordial Helium fraction\\
        $T_\text{CMB}$ & $2.7255\,\text{K}$ & CMB temperature today\\
        \hline
    \end{tabular}
    \caption{Baseline flat \textsc{Planck2018} \texttt{EE+BAO+SN} cosmological parameters used for all tests unless otherwise indicated. The dark energy density $\Omega_{\rm DE}$ is chosen so that all densities add up to critical and thus $\Omega_\text{k}=0$ serves as a constraint, not a parameter.}
    \label{tab:cosmo_params}
\end{table}

\begin{figure}
    \centering
    \includegraphics[width=0.99\textwidth]{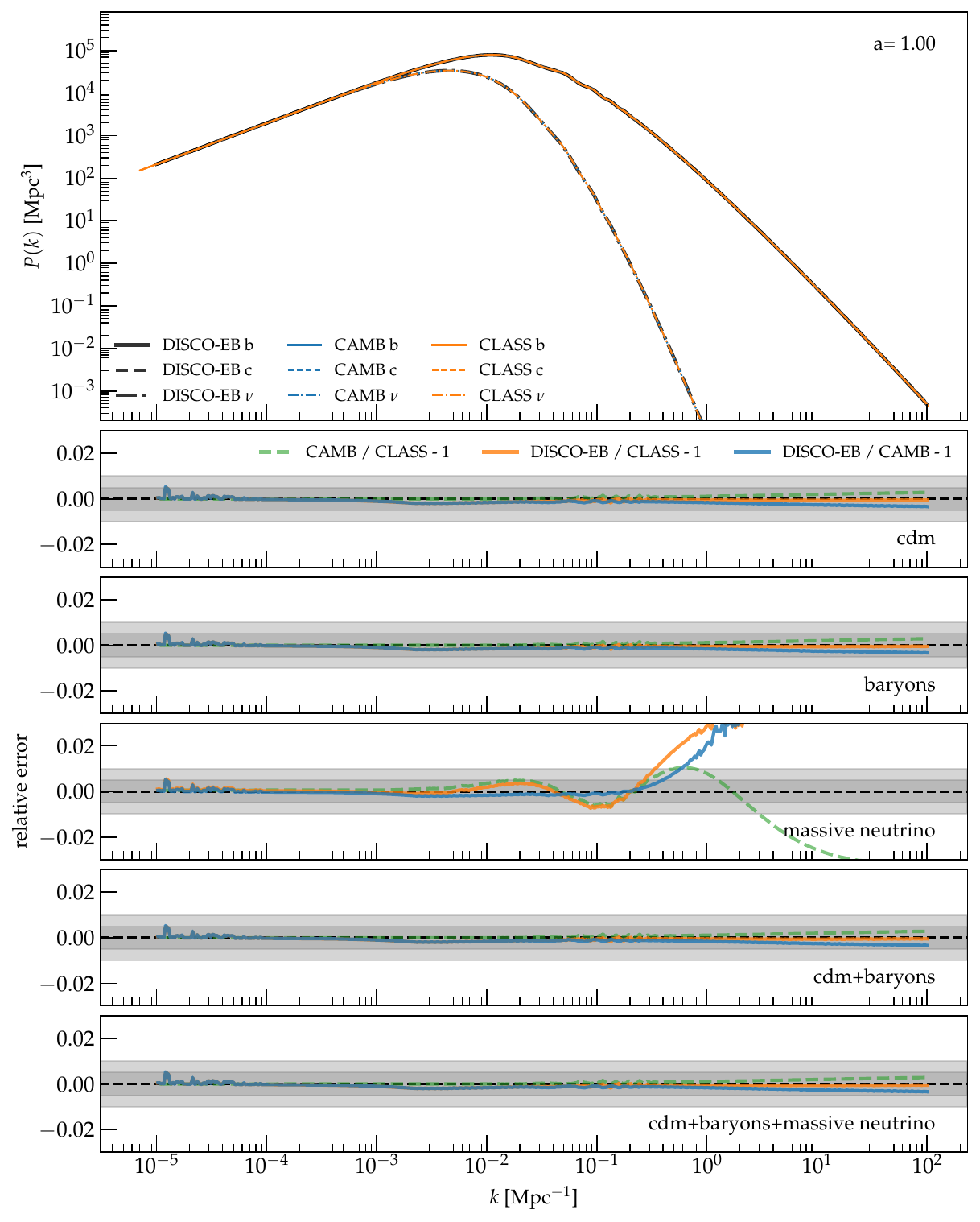}
    \caption{Comparison at $a=1$ of the performance of our differentiable Einstein-Boltzmann solver vs. \textsc{Camb} and \textsc{Class}. The top panel shows the power spectrum of density perturbations in baryons (solid), CDM (dashed), and the $0.06\,\text{eV}$ massive neutrino (dot-dashed). The lower panels indicate the relative errors $P/P_\text{ref}-1$ for CDM, baryons, massive neutrinos, CDM+baryons, and total matter = CDM+baryons+massive neutrino. The grey bands indicate 1 per cent deviation (light grey) and 0.5 per cent deviation (darker grey). All EB solvers were run on high-accuracy settings for this comparison, but we note that it might well be possible to improve agreement further with other parameter choices. }
    \label{fig:comp_CLASS_z0}
\end{figure}

\subsection{Comparison with \textsc{CAMB} and \textsc{CLASS}}
\label{sec:validation_class_camb}
In this subsection, we validate the performance of the \codename{} Einstein-Boltzmann solver against the well-established and commonly used \textsc{Camb}\footnote{Available from \url{https://camb.info} } \citep{CAMB}, specifically version 1.5.0, and \textsc{Class}\footnote{Available from \url{https://lesgourg.github.io/class_public/class.html}} \citep{CLASS}, version 3.2.0, software packages. We use both codes via their \textsc{Python} interfaces.

\begin{table}[h]
    \centering
    \begin{tabular}{l}
         \texttt{AccuracyBoost = 3} \\
         \texttt{lAccuracyBoost = 3} \\
        \texttt{DoLateRadTruncation = False} \\
        \texttt{MassiveNuMethod= 'Nu\_int'}\\
        \texttt{Transfer.high\_precision = True}\\
        \texttt{Transfer.accurate\_massive\_neutrino\_transfers = True}\\
        \texttt{Reion.Reionization = False}
    \end{tabular}
        \caption{Parameters used for the reference high-precision \textsc{Camb} runs.} 
        \label{tab:parameters_camb}
\end{table}

\begin{table}[h]
    \centering
    \begin{tabular}{l}
    \texttt{l\_max\_g} =  \texttt{64} \\
    \texttt{l\_max\_pol\_g} =  \texttt{64} \\
    \texttt{l\_max\_ur} =  \texttt{64} \\
    \texttt{l\_max\_ncdm} =  \texttt{64} \\
    \texttt{use\_ppf = no}\\
    \texttt{radiation\_streaming\_approximation = 3}\\
    \texttt{ncdm\_fluid\_approximation = 3}\\
    \texttt{ur\_fluid\_approximation = 3}\\
    \texttt{reio\_parametrization = reio\_none}
    \end{tabular}
    \caption{Parameters used for the reference high-precision \textsc{Class} runs. Note that for \codenameEB{}, we used only $\ell_\text{max}=32$.}
    \label{tab:parameters_class}
\end{table}

Since \codename{} is aimed at large-scale structure studies, we are mainly interested in the post-recombination regime. We use high accuracy settings for all three codes, listed in \autoref{tab:parameters_camb} and \autoref{tab:parameters_class}, for \textsc{Camb} and \textsc{Class} respectively, to boost the accuracy of the integration results. For \codenameEB{}, we adopted a maximum $\ell$ of the hierarchy truncation of $32$, as well as the 5-point neutrino momentum integration. Note that we found that \textsc{Class} enables by default some optimizations which can lead to relatively large differences, notably we had to disable some optimizations for massive neutrinos and the dark energy fluid.

In all tests, we employ the cosmological parameters listed in \autoref{tab:cosmo_params} as our baseline cosmological model and compute the evolution from adiabatic (isentropic) initial conditions using \codenameEB, \textsc{Camb}, and \textsc{Class}. We compare the resulting power spectra for baryons, CDM, the massive neutrino, as well as the combined total matter (CDM+baryon+massive neutrino) and the `$b+c$' (baryon+CDM) spectra to those obtained from \textsc{Camb} and \textsc{Class} at $z=0$ in \autoref{fig:comp_CLASS_z0}. 

In all cases, we find excellent agreement between the three codes, with relative differences of at most a few per mille for all quantities of interest, with the sole exception of the massive neutrino perturbations. Generally the difference is similar w.r.t. both \textsc{Class} and \textsc{Camb}, with possibly slightly better agreement between \codenameEB{} and \textsc{Class} for everything but the massive neutrinos spectrum, and slightly better agreement between \codenameEB{} and \textsc{Camb} for the massive neutrinos. We find that the spectra of massive neutrino perturbations differ at more than 1 per cent for $k\gtrsim 1\,\text{Mpc}^{-1}$, where they are suppressed by about $10^{-6}$ relative to the CDM perturbations due to free streaming. In practice, these differences are therefore not significant for the matter power spectrum, and entirely attributable to the chosen numerical precision. 

A similar comparison at $z=99$ ($a=10^{-2}$) (shown in \autoref{fig:comp_CLASS_z99} in the appendix) yields very similar results with few per-mille differences in all perturbations. Slightly larger differences exist for baryons, presumably due to differences in the recombination solver used. To improve agreement even further at higher $z$, we would therefore have to improve the quality of the recombination solver, which we leave for future work (see discussion in \autoref{sec:thermal_history}). Again, larger differences exist for massive neutrinos in the free-streaming suppressed regime, which are however of no real importance.

All in all, the agreement is therefore excellent, and we conclude that \codenameEB{} is able to reproduce the results of \textsc{Camb} and \textsc{Class} at the per mille accuracy level for all quantities of interest. We caution that the tiny differences between the results of \textsc{Camb} and \textsc{Class} might still be reduced by more optimal parameter choices.

%%%%%%%%%%%%%%%%%%%%%%%%%%%%%%%%%%%%%%%%%%%%%%%%%%%%%%%%%%%%%%%%%
\subsection{Gradient of the power spectrum w.r.t. cosmological parameters}
\begin{figure}[h!]
    \centering
    \includegraphics[width=0.99\textwidth]{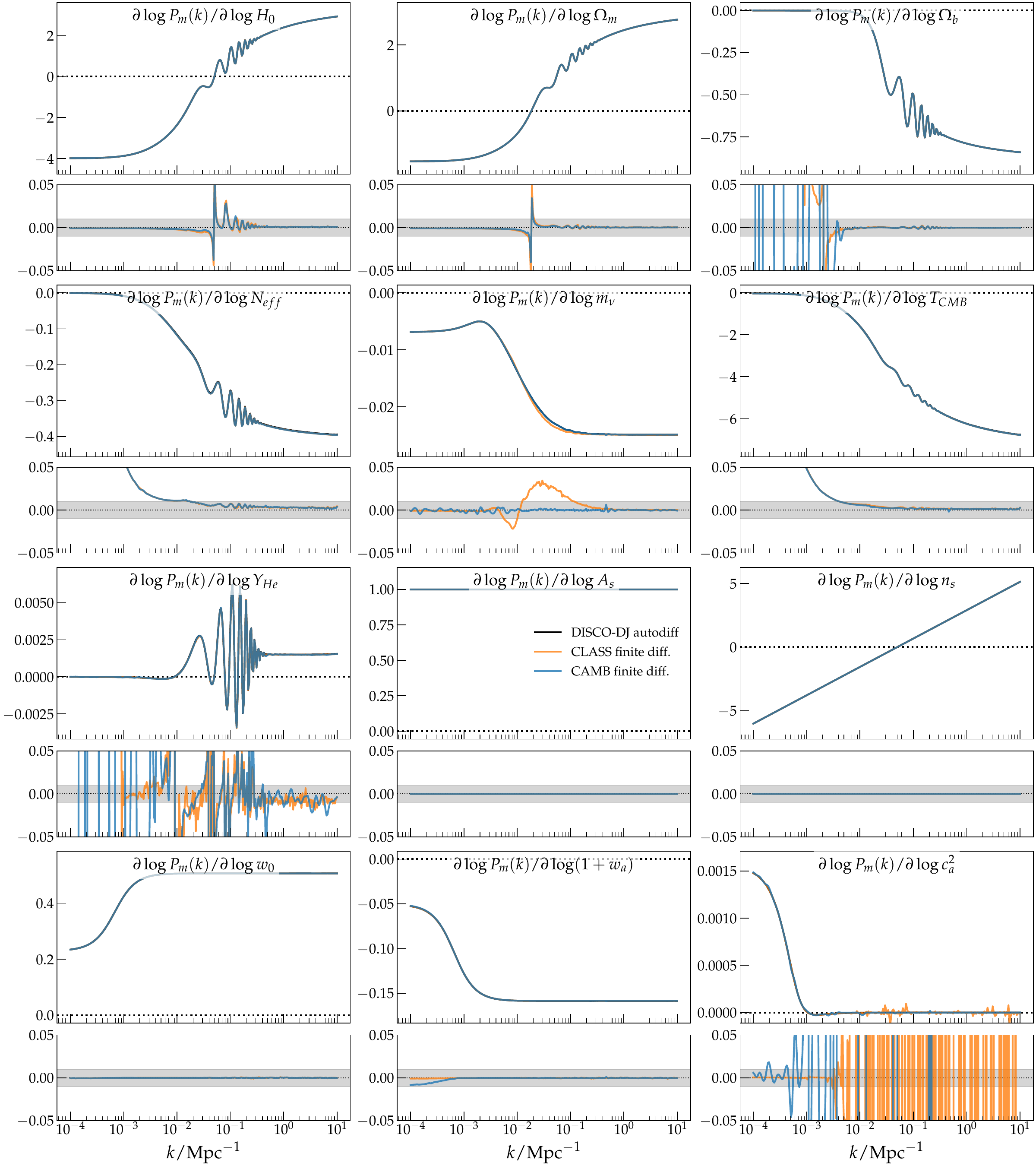}
    \caption{Derivative of the total matter (CDM+baryon+massive neutrino) density power spectrum at fixed time ($a=1$) with respect to all cosmological model input parameters, evaluated at the fiducial parameter values. We show the result of the \codenameEB{} autodiff computation (black) and a second-order finite-difference approximation based on \textsc{Class} (orange) and \textsc{Camb}, computed using eq.~\eqref{eq:fd} with $\epsilon=10^{-2}$. The respective derivative is indicated by a label in each plot. The top panel in each case represents the value of the derivative as a function of the wave number, while the bottom panel shows the relative difference between the \codenameEB{} autodiff computation and the finite difference solutions.} 
    \label{fig:P_of_k}
\end{figure}

Having a fully differentiable Einstein-Boltzmann solver permits taking derivatives with respect to any parameter of the model. In this subsection, we demonstrate this capability by computing the Jacobian matrix of the total matter (CDM+baryon+massive neutrino) density power spectrum at fixed time ($a=1$) with respect to all cosmological model input parameters. This is achieved by wrapping the power spectrum calculation in a \textsc{Jax} function, which takes the parameters as an argument, and then calls the \texttt{jacfwd} method of the \textsc{Jax} library. Specifically, let $\boldsymbol{\vartheta}\in\mathbb{R}^n$ be the vector of cosmological model input parameters, where $n=12$ for our baseline cosmology (see \autoref{tab:cosmo_params}), and specifically 
\begin{align}
    \boldsymbol{\vartheta} =\left(H_0,\Omega_m,\Omega_b,N_\text{eff},m_\nu,T_\text{CMB},Y_\text{He},A_s,n_s,w_0,w_a,c_a^2\right)^\top.
\end{align}
In principle, all physical constants and other parameters, including e.g. numerical parameters, could be added here. Let $P(k,a\mid\boldsymbol{\vartheta})$ be the vector of power spectrum values at fixed time, then we are interested in computing the logarithmic Jacobian 
\begin{align}
    \text{J}_{\boldsymbol{\vartheta}} := \boldsymbol{\nabla}_{\boldsymbol{\vartheta}} \log P(k,a\mid\boldsymbol{\vartheta}).
\end{align}
The result of the autodiff computation is shown in \autoref{fig:P_of_k} as a solid black line. The respective component of the Jacobian is indicated by a label in each plot. For comparison, we also show the finite-difference approximation of the Jacobian computed with \textsc{Camb} and \textsc{Class} (note that we have adopted $\ell_\text{max}=32$ for \textsc{Class} in this case). Specifically, we compute the finite difference approximation of the Jacobian as
\begin{align}
    \text{J}_{\vartheta_i}^\text{FD} = \frac{\log P(k,a\mid\boldsymbol{\vartheta}+\tfrac{\epsilon}{2}\vartheta_i\boldsymbol{e}_i) - \log P(k,a\mid\boldsymbol{\vartheta}-\tfrac{\epsilon}{2}\vartheta_i\boldsymbol{e}_i)}{\epsilon \vartheta_i} + \mathcal{O}(\epsilon^2),\label{eq:fd}
\end{align}
which means that for second-order finite differences two calls to the EB solver are needed per model parameter. We show the results for $\epsilon=10^{-2}$ in \autoref{fig:P_of_k}. The top panel in each case represents the value of the derivative as a function of the wave number, while the bottom panel shows the relative difference between the \codenameEB{} autodiff computation and the finite difference solutions using \textsc{Camb} and \textsc{Class}.

We note that the autodiff evaluation and the finite difference approximation agree generally very well to mostly better than one per cent. We found that a good bias-variance tradeoff requires $\epsilon \approx 10^{-2}$, while for smaller values, the finite difference estimates become very noisy. Nonetheless, where visible, the autodiff version shows fewer signs of numerical artefacts. The only significant difference is found for the neutrino mass derivative w.r.t. the finite difference result for \textsc{Class}, which is attributable to the reduced accuracy we used (specifically not setting \texttt{ncdm\_fluid\_approximation} as this leads to prohibitive runtimes). We note that the finite difference approximation is computationally expensive, since it requires two calls to the EB solver per parameter, while the autodiff computation is computationally much cheaper. Due to the large impact of optimizations and specific parameter choices on the performance of the EB solvers, we prefer to refrain from a detailed comparison of the computational performance of the autodiff and finite difference approaches.

%%%%%%%%%%%%%%%%%%%%%%%%%%%%%%%%%%%%%%%%%%%%%%%%%%%%%%%%%%%%%%%%%
\begin{figure}
    \centering
    \includegraphics[width=0.99\textwidth]{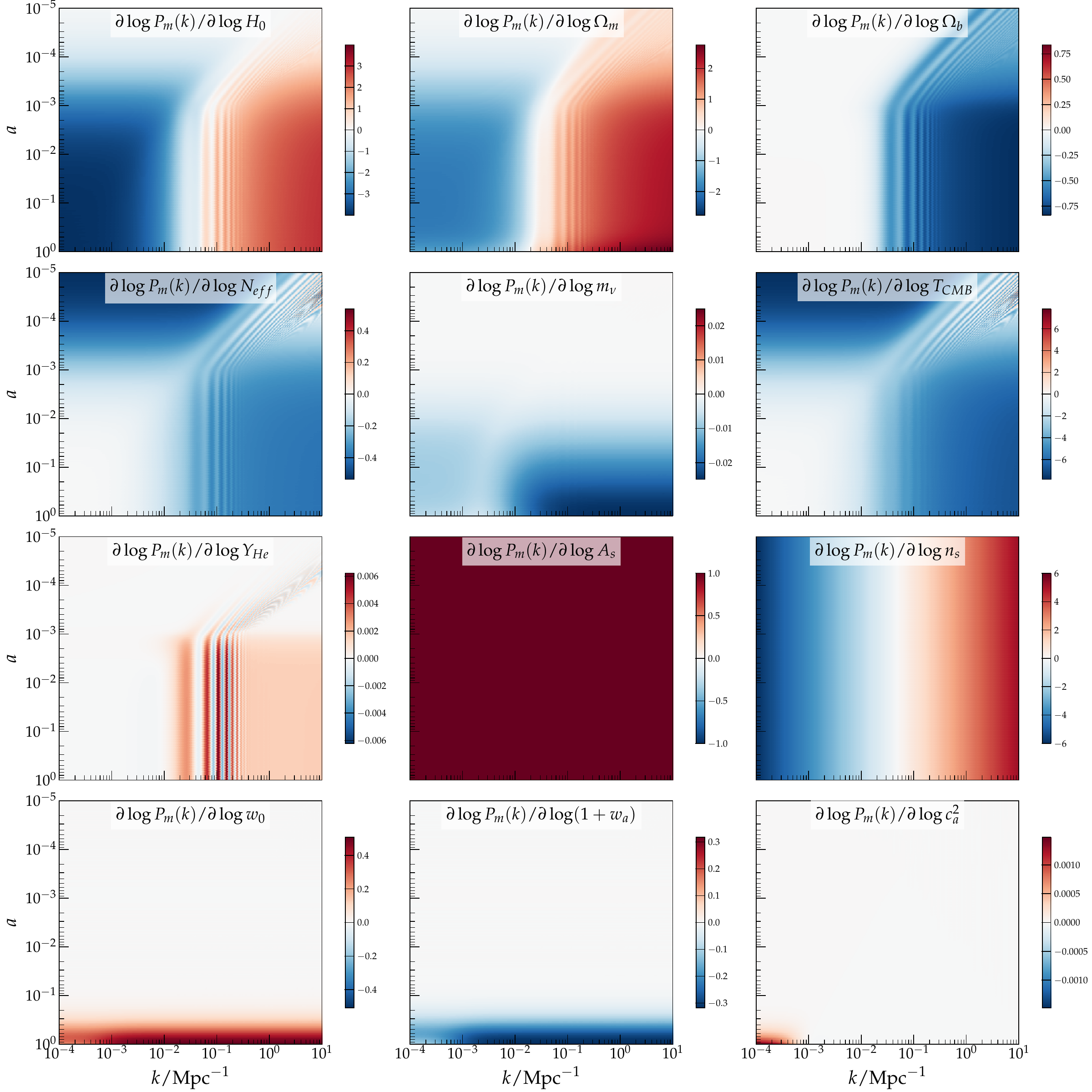}
    \caption{Time and spatial dependence of the logarithmic derivative of the total matter power spectrum w.r.t. the cosmological parameters. Each panel shows the colour-coded change in the power spectrum in arbitrary normalisation, with red indicating an increase and blue a decrease w.r.t. an arbitrary baseline, as a function of scale factor ($y$-axis) and wave number ($x$-axis). Compare with \autoref{fig:P_of_k} which represents a slice at fixed time $a=1$.}
    \label{fig:P_of_k_and_a}
\end{figure}

Finally, in \autoref{fig:P_of_k_and_a}, we show full time and space dependence of the derivative of the total matter power spectrum (cdm+baryons+massive neutrino) w.r.t. the model parameters. Each panel shows the colour-coded change in the power spectrum, with red colours indicating an increase and blue a decrease, as a function of scale factor ($y$-axis) and wave number ($x$-axis). A white colour indicates that the respective parameter has no impact on the matter power spectrum at that time and scale. For this analysis, we extend this (qualitative) analysis to earlier times ($a=10^{-5}$) revealing how the BAO feature propagates to increasingly larger scale prior to decoupling at $a\approx 10^{-3}$. Also intuitively, the neutrino mass dependence clearly shows when the neutrino becomes non-relativistic at $z\lesssim 100$, while the signature of the dark energy EOS parameters $w_0$, $w_a$ and $c_a^2$ is restricted to very late times $z\lesssim 1$, with $c_a^2$ clearly impacting only the sound horizon in the dark energy fluid.

%%%%%%%%%%%%%%%%%%%%%%%%%%%%%%%%%%%%%%%%%%%%%%%%%%%%%%%%%%%%%%%%%
\section{Applications}
In this section, we demonstrate a few applications of the differentiable Einstein-Boltzmann solver. We note that these are not meant to be exhaustive in any way. Specifically, two main applications of a differentiable solver will be for parameter inference and the construction of emulators, aspects that we postpone to future work. Nonetheless, the cases we highlight here are meant to demonstrate the applicability and integration of the solver for a few select applications relevant to large-scale structure cosmology.
\label{sec:applications}
\subsection{Redshift-space correlators for large-scale structure studies}
The matter power spectrum itself is not directly observable but can be constrained e.g. through the clustering of galaxies. Galaxies are, however, biased tracers of the underlying matter distribution (cf. \cite{Desjacques:2018biasreview} for a review) and their three-dimensional distribution can generally only be determined in redshift space \cite{Kaiser:1987}. As a next proof-of-concept test, we demonstrate a differentiable calculation of the redshift-space power spectrum and the correlation function multipoles of biased tracers, as this is a common input for large-scale structure studies.

\subsubsection{Linear redshift space distortions}
The proper motions $v_\parallel$ of galaxies parallel to the line-of-sight (LOS) contribute to the measured redshift 
\begin{align} 
    1+z_\text{obs} = (1+z_\text{cosmo})\;(1+v_\parallel/c)
\end{align}
and lead to the redshift space distortion (RSD) effect.  At linear order and in the Newtonian limit, the redshift-space power spectrum can be computed from the matter power spectrum using the Kaiser formula \cite{Kaiser:1987}. For linearly biased tracers with density contrast $\delta_g$ it states that $ \delta_g^\text{Kaiser} :=  b\, \delta_m - \mu^2 \,\frac{\theta_m}{\mathcal{H}}$ and so
\begin{align}
     P_\text{lin}(k,\mu;\,z)=  \left[b(z)\, \delta_m(k,z) - \frac{\mu^2}{\mathcal{H}(z)} \,\theta_m(k,z)\right]^2\label{eq:kaiser}
\end{align}
where $b$ is the linear bias coefficient (which we fix to $b=2$ here), $\mu:=\cos\alpha$ for an angle $\alpha$ between $\boldsymbol{k}$ and the line-of-sight, $\theta_m$ is the matter velocity divergence, and $\mathcal{H}=a'/a$ is the conformal-time expansion rate. Further, one needs to assume an irrotational velocity field ($\boldsymbol{v} = \boldsymbol{\nabla} \nabla^{-2}\theta$) and the absence of velocity bias (i.e. galaxy velocity $\theta_g = \theta_m$) for this relation to hold. Commonly, the growth rate $f$ is used to express $\theta_m$ in terms of $\delta_m$, but since we have both fields from the EB solver, it is more convenient to work directly with the velocity divergence. The fields $\delta_m$ and $\theta_m$ are readily obtained from \codenameEB{}.

%%%%%%%%%%%%%%%%%%%%%%%%%%%%%%%%%%%%%%%%%%%%%%%%%%%%%%%%%%%%%%%%%
\subsubsection{Correlation function multipoles}

\begin{figure}
    \centering
    \includegraphics[width=0.99\textwidth]{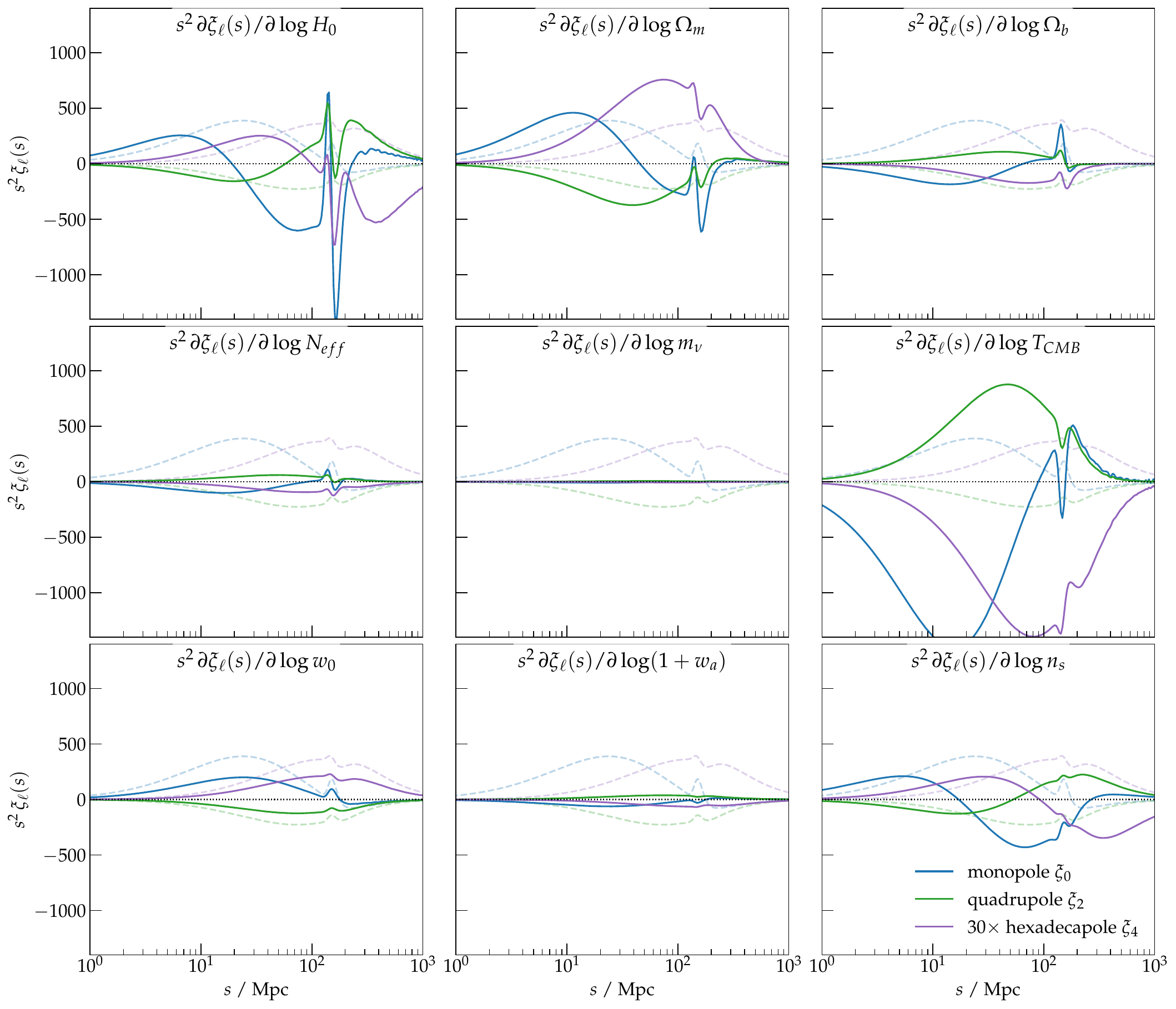}
    \caption{Derivative of the redshift-space correlation function multipoles w.r.t. the cosmological parameters. Each panel shows the derivative as a function of scale $r$ of the monopole $\xi_0(r)$ (blue), the quadrupole $\xi_2(r)$ (green), and the hexadecapole $\xi_4(r)$ (purple), the respective correlation function multipoles (without derivative applied) are shown as light dashed lines. We have multiplied the hexadecapole $\xi_4$ by a factor of 30 for visual clarity. In this case the derivative is carried through the \textsc{FFTLog} method and the Einstein-Boltzmann solver.}
    \label{fig:xi_of_r}
\end{figure}

Redshift space distortions on large scales are conveniently expressed in terms of a multipole expansion of the power spectrum. These redshift-space power spectrum multipoles are commonly defined as the following expansion in Legendre polynomials \cite{Cole:1994rsdmultipole,Hamilton:2000}
\begin{align}
    P_\ell(k;\,z) := \frac{2\ell+1}{2} \int_{-1}^1 {\rm d}\mu \;P(k,\mu;\,z)\; L_\ell(\mu)\;,
\end{align}
where $\mu=\cos\alpha$ is the cosine of the angle $\alpha$ between the wave vector $\mathbf{k}$ and the line-of-sight direction $\mathbf{n}$, $P(k,\mu;\,z)$ is the anisotropic power spectrum of the tracer, and $L_\ell$ is the Legendre polynomial of order $\ell=0,1,2,\dots$.  Then the redshift-space power spectrum multipoles for the Kaiser formula (eq.~\eqref{eq:kaiser}, \cite{Kaiser:1987}) are given by
\begin{subequations}
\begin{align}
    P_\ell(k;\,z) &= \frac{2\ell+1}{2} \int_{-1}^{+1}\text{d}\mu\;\left( b(z)\, \delta_m(k;\,z) -  \frac{\mu^2}{\mathcal{H}(z)}\theta_m(k;\,z) \right)^2\;L_\ell(\mu) 
    \intertext{and therefore with $P_{\delta\delta}:=\delta_m \delta_m^\ast=P_m$, $P_{\delta\theta}:=\delta_m\theta_m^\ast$, and $P_{\theta\theta}:=\theta_m\theta_m^\ast$ the only non-zero multipoles for the linear Kaiser model are}
    P_0(k;\,z) &= b^2 P_{\delta\delta} -\frac{2 b}{3 \mathcal{H}} P_{\delta\theta} + \frac{1}{5\mathcal{H}^2} P_{\theta\theta} \\
    P_2(k;\,z) &= -\frac{4 b}{3 \mathcal{H}} P_{\delta\theta} + \frac{4}{7\mathcal{H}^2} P_{\theta\theta} \\
    P_4(k;\,z) &= \frac{8}{35 \mathcal{H}^2} P_{\theta\theta}\;.
\end{align}
\end{subequations}
Finally, the correlation function multipoles are defined in terms of the power spectrum multipoles via the integral
\begin{align}
    \xi_\ell(r;\,z) = \frac{\text{i}^\ell}{2\pi^2} \int_0^\infty \text{d}k\, k^2 \;P_\ell(k;\,z) \; j_\ell(kr)\;,
\end{align}
where $j_\ell$ is the spherical Bessel function of order $\ell$. These integrals are typically efficiently computed using the \textsc{FFTLog} method \cite{Talman:1978,Hamilton:2000}, which can be  implemented in \textsc{Jax} in a few lines, and thus allows differentiable computation of the correlation function. 

We show the resulting derivatives of the monopole, quadrupole, and hexadecapole w.r.t. the baseline cosmological parameters in \autoref{fig:xi_of_r}. The respective derivative is indicated by a label in each plot. We have omitted the dark energy parameters here as they predominantly affect the growth history and thus only the normalisation of the correlation function multipoles. Clearly visible are the BAO features in many derivatives (but not all). The calculation in this case was carried out using 512 modes geometrically distributed between $k=10^{-5} \,\text{Mpc}^{-1}$ and $k=100 \,\text{Mpc}^{-1}$ since the \textsc{FFTlog} transform is relatively noisy with fewer sampling points, and even in this case, some ringing can be observed at $r\gtrsim 200 \,{\rm Mpc}$. This could likely be improved by additional tweaks to the \textsc{FFTlog} method, which goes beyond our minimal \textsc{FFTlog} implementation, and which we leave for future work.

\subsubsection{Two-dimensional power spectrum}

In a more realistic analysis, in addition to the Kaiser effect, typically also the Finger-of-God effect and the Alcock-Paczynski effect \cite{AP:1979} are considered. 

\paragraph{The Alcock-Paczynski (AP) effect.} It arises from the fact that the observables redshift and angle on the sky need to be converted to length scales assuming a fiducial cosmology which might not be the true cosmology. The AP effect can be expressed as a modification of the observed wave vector $k_\text{obs}$ and the observed angle $\mu_\text{obs}$ in terms of the true wave vector $k$ and angle $\mu$ as
\begin{align}
    k_\text{obs}(k,\mu,z) &:= \frac{k}{q_\perp}\left[ 1+ \mu^2\left(\frac{q_\perp^2}{q_\parallel^2}-1\right)\right]^\frac{1}{2}, &
    \mu_\text{obs}(\mu,z) &:= \mu\frac{q_\perp}{q_\parallel} \left[1+\mu^2\left(\frac{q_\perp^2}{q_\parallel^2}-1\right)\right]^{-\frac{1}{2}}
\end{align}
and volume deformation $V_\text{obs} := q_\perp^2 q_\parallel$. The deformation parameters are 
\begin{align}
    q_\perp(z) &:= \frac{D_A(z)}{D^\text{fid}_{A}(z)} & q_\parallel(z) &:= \frac{\mathcal{H}^{\text{fid}}(z)}{\mathcal{H}(z)}
\end{align}
where $D_A(z)$ is the angular diameter distance and $\mathcal{H}(z)=a'/a$ is the conformal Hubble function. The superscript ${}^\text{fid}$ indicates the value in the reference (fiducial) cosmology.
which leads to a volume deformation $V_\text{obs} := q_\perp^2 q_\parallel$.

\paragraph{The Finger-of-God (FoG) effect.} It arises from the fact that galaxies in clusters have a velocity dispersion which leads to a smearing of the observed power spectrum. The effect can be modelled as a Lorentzian damping of the power spectrum in redshift space \cite{Percival:2004,EuclidForecast,Casas:2024}. It amounts to a multiplicative factor of the form
\begin{align}
    F_\text{FoG} &:= \frac{1}{1+k^2\mu^2 \sigma_\theta^2(z)} &\text{where}\quad \sigma_\theta^2(z) := \frac{1}{6\pi^2(\mathcal{H}^\text{fid})^2}\int P_{\theta\theta}^\text{fid}(k,z)\,\dd k 
\end{align}

\paragraph{The observed power spectrum.} Putting all together, we can write the observed (linear) power spectrum including all mentioned effects as
\begin{align}
    P_\text{obs}(k,\mu;\,z) = \frac{1}{q_\perp^2(z)\, q_\parallel(z)}\;\frac{P_\text{lin}\left(k_\text{obs}\left(k,\mu,z\right),\,\mu_\text{obs}(\mu,z);\,z\right)}{1+k_\text{obs}^2(k,\mu,z)\;\mu^2_\text{obs}(\mu,z)\;\sigma_\theta^2(z)}\;. \label{eq:observed_power}
\end{align}
Note that the power spectrum of course always also depends on all physical/cosmological parameters $\boldsymbol{\vartheta}$, which we generally have suppressed in this section to avoid cluttering the notation further.

\begin{figure}
    \centering
    \includegraphics[width=0.99\textwidth]{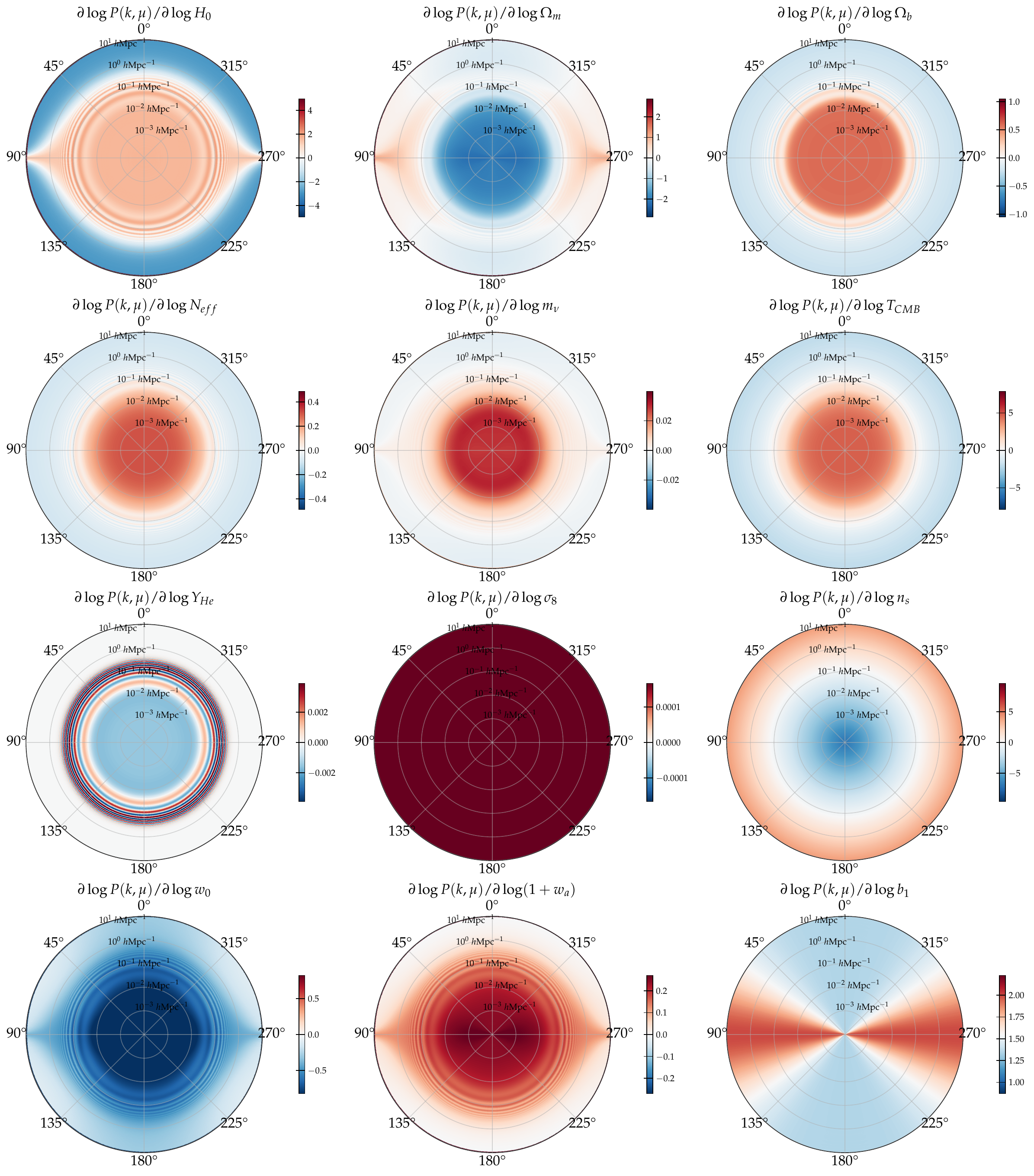}
    \caption{Logarithmic derivatives of the two-dimensional redshift-space matter power spectrum w.r.t. model parameters $\boldsymbol{\vartheta}$, $\partial\log P_\text{obs}(k,\mu)/\partial \log \vartheta_i$.  The angular coordinate in each plot is the angle between a mode $\boldsymbol{k}$ and the line-of-sight, while the radial coordinate is $k$, which is clipped to be between $k_\text{min}=10^{-4}\;\text{Mpc}^{-1}$ (the center point) and $k_\text{max}=10\;\text{Mpc}^{-1}$ (the outer circle). The colour scale indicates a relative increase (red) vs. decrease (blue). }
    \label{fig:observed_pk_alldiff}
\end{figure}

In \autoref{fig:observed_pk_alldiff}, we show the logarithmic gradients of the observed matter power spectrum $\partial\log P_{\rm obs}/\partial\log \vartheta_i$ with respect to a key subset of model parameters $\boldsymbol{\vartheta}$ as obtained with \codename{}. Wave vectors parallel to the LOS are indicated in the $0^\circ$ direction, while modes perpendicular are at $90^\circ$ and $270^\circ$. The radial coordinate is $\log k$, which is clipped to be between $k_\text{min}=10^{-4}\;\text{Mpc}^{-1}$ (the center point) and $k_\text{max}=10\;\text{Mpc}^{-1}$ (the outer circle). The colour scale indicates a positive dependence of $P_{\rm obs}$ on the parameter (red) or a negative dependence (blue). A white region indicates that the information content due to parameter $\vartheta_i$ of the power spectrum vanishes. 

A few aspects are notable. First, the BAO feature is clearly visible in the derivatives w.r.t. $H_0$, $Y_{\rm He}$, and the DE EOS parameters $w_0$ and $w_a$, indicating that these parameters have a strong impact on the BAO feature (i.e. causing a shift in the BAO scale). Thirdly, the derivatives w.r.t. $\Omega_m$ and the bias parameter $b$ reveal a quadrupole pattern in the $k$-$\mu$ plane which is due to their impact on the redshift space distortion effect ($\Omega_m$ determines the growth rate $f={\rm d \log D_+}/{\rm d}\log a$, and $b$ and $f$ directly determine the RSD effect in the Kaiser formula).

%%%%%%%%%%%%%%%%%%%%%%%%%%%%%%%%%%%%%%%%%%%%%%%%%%%%%%%%%%%%%%%%%
\subsection{Fisher-forecasting with differentiable solvers}

As a proof-of-concept, we demonstrate an application to compute the Fisher information matrix for forecasting. Fisher forecasts are widely used in cosmology for experimental design \citep{Tegmark1997, Amendola2013}, and they require computing derivatives of the likelihood with respect to the cosmological parameters. Typically, these derivatives are computed with finite differences, which requires careful tuning for convergence \cite{Bhandari2021}. Autodiff makes the computation of the Fisher matrix easier and avoids the convergence problem. 

We demonstrate the computation of a Fisher forecast for the `optimistic' spectroscopic Euclid survey with 4 bins as described in \cite{EuclidForecast,Casas:2024}. For a Gaussian likelihood, the elements of the Fisher matrix for a summary statistic mean $y(\ell)$ are
\begin{align}
    F_{ij} = \sum_\ell \frac{\partial y(\ell)^\top}{\partial \vartheta_i} C^{-1}(\ell) \frac{\partial y (\ell)}{\partial \vartheta_j}
\end{align}
 where $C(\ell)$ is the covariance matrix computed at the fiducial cosmology. Following \cite{EuclidForecast}, we account for effective bias, anisotropies due to redshift space distortions, and redshift uncertainty in the observed two-dimensional power spectra, given by eq. \eqref{eq:observed_power}. When considering a tomographic survey, the bias parameter $b_1(z)$ and redshift error, $\sigma_z(z)$ are of course redshift dependent. Specifically, one has that redshift errors grow roughly like $\sigma_z = (1+z) \sigma_{z,0}$.

\paragraph{Linear covariance and survey properties.} In our simplified proof-of-concept analysis, we assumed the linear approximation of the covariance matrix \cite{EuclidForecast,Casas:2024}:
\begin{align}
    C(\boldsymbol{k}, \boldsymbol{k'}) \approx \frac{2 (2\pi)^2}{V_s(z)}\;P_\mathrm{obs}^2(\|\boldsymbol{k}\|, \mu; z)\; \delta_D(\boldsymbol{k} - \boldsymbol{k'}),
\end{align}
where $\delta_D$ is the Dirac delta distribution and $V_s(z)$ is the survey volume for the redshift bin at $z$. We assume a galaxy density for this bin of 
\begin{align}
    n(z) = \frac{\dd^2 N(\Omega, z)}{\dd\Omega \dd z} \frac{A_\mathrm{survey}}{V_s(z)}\Delta z.
\end{align}
The survey area in the sky is $A_\mathrm{survey} = 15000$ deg$^2$, and the rest of the parameters are set as in Table~3 in Ref. \cite{EuclidForecast} (equivalent to Table~3 in \cite{Casas:2024}).

\paragraph{Observed power spectrum.} In the standard Euclid forecasting \cite{EuclidForecast,Casas:2024}, the observed power spectrum $\hat{P}_\mathrm{obs}$ is the linear spectrum $P_\text{obs}$ from eq.~\eqref{eq:observed_power} modified in three additional ways: (1) the damping of the BAO feature due to non-linearities is mimicked by a blending of the linear spectrum with a filtered version with BAO wiggles removed. Specifically, as in \cite{Casas:2024}, we apply a 3rd order Savitzky-Golay filter of width $\Delta\log k\approx 1.53$ to obtain the filtered $P_\text{obs,nw}$. This is then blended with the original spectrum to yield
\begin{align}
    \tilde{P}_\text{obs}(k,\mu;\,z) &:= P_\text{obs}(k,\mu;\,z) \;e^{-g(k,\mu;\,z) k^2} + P_\text{obs,nw}(k,\mu;\,z) \;\left(1-e^{-g(k,\mu;\,z) k^2}\right) 
    \intertext{with the transition scale}
    g(k,\mu;\,z) &:= \frac{1}{6\pi^2}\left[1-\mu^2+\mu^2\left(1+f^\text{fid}(k,z)\right)^2\right]\;\int\dd k\;P_{\delta\delta}(k,z)\;.
\end{align}
Then, (2) the result is multiplied by a Gaussian dispersion kernel (cf. \cite{PeacockDodds:1994,ChavesMontero:2018}) modelling the impact of redshift measurement errors, and (3) receives an additional shot-noise contribution (note that we follow here in notation \cite{Casas:2024}, rather than \cite{EuclidForecast} who absorb the $1/n(z)$ contribution into an effective survey volume)
\begin{align}
    \hat{P}_\text{obs}(k,\mu;\,z) &:= \tilde{P}_\text{obs}(k,\mu;\,z)\; G_z(k,\mu;\,z) + \left[\frac{1}{n(z)} + \delta P_s(z)\right]
    \intertext{where $\delta P_s(z) = \delta P_s^\text{bin}$ is a shot-noise nuisance parameter for each bin (with fiducial value 0), and the redshift error kernel is given by}
    G_z(k,\mu;\,z)&:= \exp\left(-\frac{c^2k^2}{(\mathcal{H}^\text{fid})^2} \mu^2 \sigma_{z,0}^2 \right) \;,
\end{align}
where $ \sigma_{z,0}(1+z)=\sigma_z$ is the measurement error on the redshift (we adopt $\sigma_{z,0}=0.001$ as the fiducial value for the spectroscopic Euclid survey). Similarly to \cite{Casas:2024} we allow for an additional nuisance parameter correcting the bias, i.e. we replace in eq.~\eqref{eq:observed_power}
\begin{align}
    b(z) \to b(z) +  \frac{\sigma_8(0)}{\sigma_8(z)}\;\delta b(z)\;,  
\end{align}
where $b(z) = b^\text{bin}$ is the fiducial mean bias per bin and $\delta b(z) = \delta b^\text{bin}$ is an additional nuisance parameter for which the trivial dependence on changes in the normalisation is corrected by the scaling with $\sigma_8$. 

\begin{figure}
    \centering
    \includegraphics[width=0.99\textwidth]{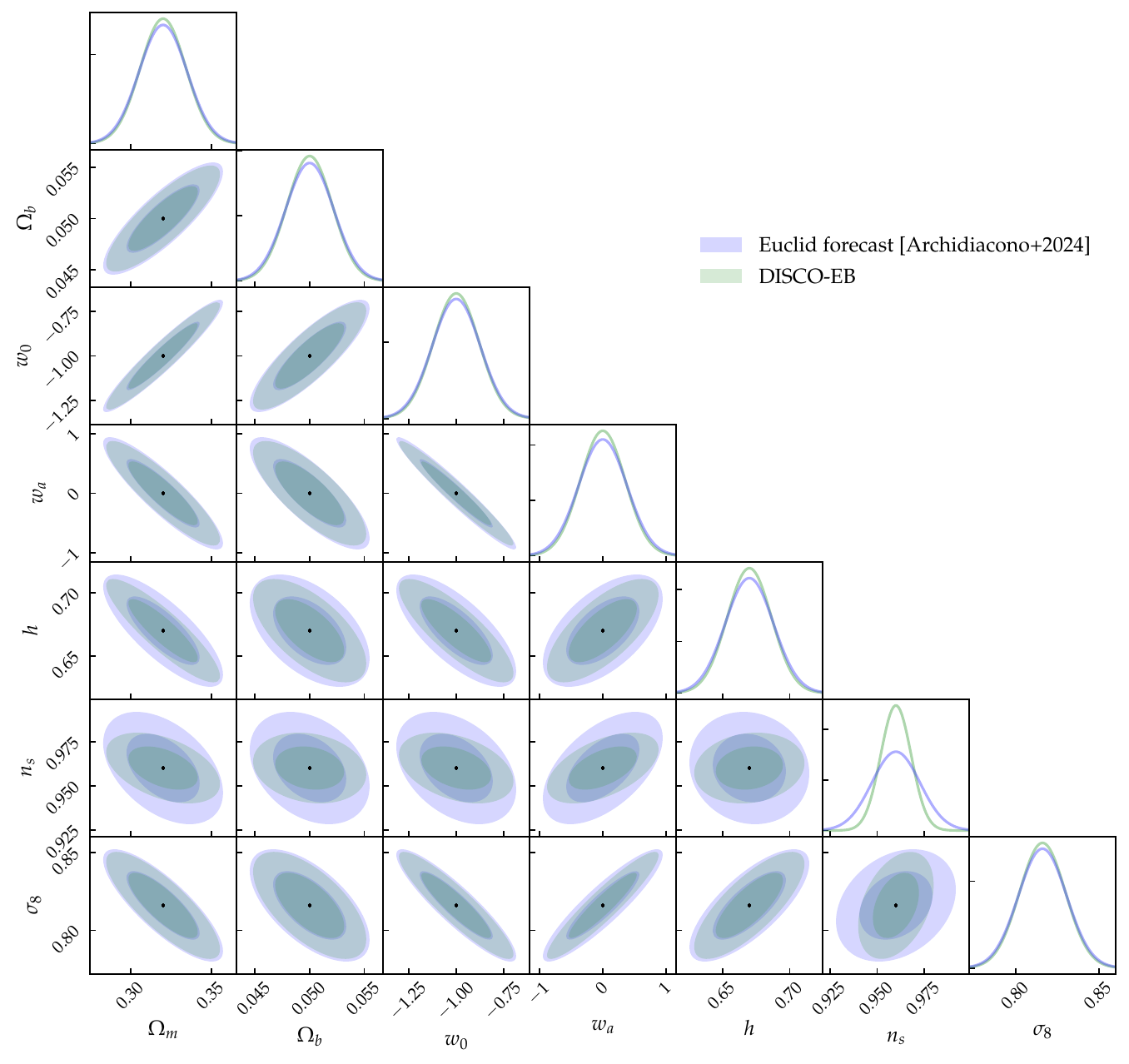}
    \caption{$1-$ and $2-\sigma$ Fisher contours predicted for the `optimistic' Euclid spectroscopic galaxy clustering survey using a simplified model based on \codenameEB{} (green) in comparison to comparable predictions from the official Euclid forecast \citep[][blue]{Archidiacono:2024}.}
    \label{fig:Fisher}
\end{figure} 

\paragraph{Fisher matrix.} The elements of the Fisher matrix are then given by \cite{Tegmark1997, Seo2003, EuclidForecast} 
\begin{align}
    F^\mathrm{bin}_{ij} := \frac{1}{8\pi^2} \int^1_{-1} \dd\mu \int^{k_\mathrm{max}}_{k_\mathrm{min}} \dd k\,k^2\,  \frac{\partial \log \hat{P}_\mathrm{obs}}{\partial \vartheta_i} \;\frac{\partial \log \hat{P}_\mathrm{obs} }{\partial \vartheta_j} \;V_s(z_\mathrm{bin})
\end{align}
for each tomographic bin. In the k-space integral, we employ Euclid's `optimistic' cuts on the wave number $k$ included in the Fisher analysis as $k_\text{min}=10^{-3}\;h^{\text{fid}}\text{Mpc}^{-1}$ and $k_\text{max}=0.3\;h^\text{fid}\text{Mpc}^{-1}$.  We assume the four tomographic bins described in Table~3 of \cite{EuclidForecast}. The total Fisher matrix is then computed by summing over the different tomographic bins
\begin{equation}
    F_{ij} = \sum_\mathrm{bin} F^\mathrm{bin}_{ij},
\end{equation}
where we 
computed the power spectra for each tomographic bin at the bin's mean redshift. For this Fisher forecast, we use the Euclid fiducial values of cosmological parameters from \cite{Casas:2024}. Figure \ref{fig:Fisher} shows the resulting Fisher contours. The \codenameEB{} Fisher contours (in green) agree well with the forecast of Archidiacono et al. (2024, in prep), whose results we reproduce here, showing that \codename{} can provide reliable Fisher matrices. We see a discrepancy in expected constraints on $n_s$ in comparison to the official forecast, which we have not managed to track down and that are likely due to the minor differences in the modelling.

In addition, the Fisher matrix provided by \codename{} is itself differentiable, which allows for optimisation of the figure of merit and also for survey optimisation (see also \cite{campagne2023jax}). The source code for our forecasting will be made available along with the main source code of \codenameEB{}.

%%%%%%%%%%%%%%%%%%%%%%%%%%%%%%%%%%%%%%%%%%%%%%%%%%%%%%%%%%%%%%%%%
\section{Summary and Conclusions}
\label{sec:conclusions}

In this article, we have presented the Einstein-Boltzmann module of \codename{} (\textbf{DI}fferentiable \textbf{S}imulations for \textbf{CO}smology -- \textbf{D}one with \textbf{J}\textsc{ax}), which implements a differentiable Einstein-Boltzmann solver for cosmological perturbations and will be part of a comprehensive differentiable package for large-scale structure cosmology. \codename{} is based on the \textsc{Jax} library, which provides automatic differentiation of \textsc{Python} and \textsc{NumPy} code. Automatic differentiation is a powerful tool for the development of differentiable predictions, as it allows the efficient and accurate computation of derivatives of a model with respect to any input parameter. 

In this article, we have introduced the \codename{} Einstein-Boltzmann solver and demonstrated its performance by comparing it to the well-established \textsc{Camb} and \textsc{Class} solvers. Notably, we find that all three solvers typically agree to better than a few per mille when run with high precision settings. We have also demonstrated the accuracy of the computation of the Jacobian matrix of the total matter density power spectrum at fixed time with respect to all cosmological model input parameters by comparing it to finite-difference approximations obtained with \textsc{Camb} and \textsc{Class}. We have not (yet) provided a detailed comparison of the computational performance of the autodiff and finite difference approaches, since this would require an optimization of parameters and settings for each code given a desired target accuracy, which is beyond the scope of this first paper.

In Section \ref{sec:applications}, we have demonstrated the power of differentiable solvers by computing the Jacobian matrix of the total matter density power spectrum at fixed time with respect to all cosmological model input parameters. We have also shown the full time and space dependence of the derivative of the total matter power spectrum w.r.t. the model parameters. We have validated the correctness of the Jacobian matrix by comparing it to finite-difference approximations.
We have demonstrated the power of differentiable solvers by computing the Jacobian matrix of the total matter density power spectrum at fixed time with respect to all cosmological model input parameters. We have also shown the full time and space dependence of the derivative of the total matter power spectrum w.r.t. the model parameters. We have validated the correctness of the Jacobian matrix by comparing it to finite-difference approximations. 

The output of an Einstein-Boltzmann solver is typically the input for predictions of the summary statistic of an observable. To demonstrate how easy it is to compute such predictions with differentiable solvers, we have computed the redshift-space power spectrum and the correlation function multipoles of biased tracers. We have also demonstrated the computation of a Fisher forecast for the Euclid spectroscopic survey. While state-of-the-art modelling would have to take into account also non-linearities and other effects, these results serve as proof-of-concept use cases for our differentiable Einstein-Boltzmann solver. 

By visualisation of parameter gradients of a model, we see value also of differentiable solvers for educational purposes as they are capable of directly showing the impact of changes in the (cosmological) model parameters on the observables.

Beyond the applications shown here, we believe that a differentiable Einstein-Boltzmann solver will be useful for more efficient likelihood maximization and sampling, the development of likelihood-free inference methods, and in particular the development of differentiable non-linear models, such as differentiable $N$-body solvers, or differentiable non-linear perturbation theory, based on e.g. the effective field theory of LSS (see e.g. \cite{Ivanov:2022} for a review) or high-order Lagrangian perturbation theory (e.g. \cite{Rampf:2021sc, Schmidt:2021}). We expect that they also allow the development of more efficient emulators based on Einstein-Boltzmann solvers \cite{Arico:2021baccoemu,Bonici:2022emu,Spurio:2022,Donald-Mccann:2022}, as they allow inclusion of information from the derivatives of the power spectrum with respect to the cosmological model parameters when building the emulator and should thus require much fewer training points. Naturally, the gradient information can be made part of the emulator itself (such as \cite{Bonici:2023diffemu}), which would allow for usage of gradient information in the inference process.

\codename{} is a work in progress, and we plan to extend it in several directions. Most importantly, this module will serve as the starting point for a comprehensive differentiable package for large-scale structure cosmology, which will include a differentiable $n$-th order Lagrangian perturbation theory (LPT) module, as well as an $N$-body code. Several papers in this direction are currently in preparation. We also invite the community to extend the Einstein-Boltzmann  solver to include non-standard cosmological models, such as modified gravity and alternative dark energy and dark matter models.

We make the \codename{} code publicly available upon publication of this paper under the URL \url{https://github.com/ohahn/DISCO-EB}.
%%%%%%%%%%%%%%%%%%%%%%%%%%%%%%%%%%%%%%%%%%%%%%%%%%%%%%%%%%%%%%%%%

\acknowledgments
The authors thank Raul Angulo, Mateja Gosenca, Julien Lesgourgues, Antony Lewis, Thomas Montandon, Cristiano Porciani, and Cornelius Rampf for discussions and/or comments on an earlier draft of this paper, as well as our anonymous referee for insightful comments. We thank in particular also Julien Lesgourgues and Santiago Casas for discussions about and the sharing of their most recent Fisher forecasts for Euclid with us. The computational results presented have been achieved in part using the Vienna Scientific Cluster (VSC). The authors declare no conflicts of interest or external support in the preparation of this manuscript. 

\bibliographystyle{JHEP}
\bibliography{biblio}

\providecommand{\href}[2]{#2}\begingroup\raggedright\begin{thebibliography}{10}

\bibitem{Lifshitz:1946}
E.M.~{Lifshitz}, \emph{{On the gravitational stability of the expanding universe}}, {\emph{Zhurnal Eksperimentalnoi i Teoreticheskoi Fiziki} {\bfseries 16} (1946) 587}.

\bibitem{LifshitzKhalatnikov:1963}
E.M.~{Lifshitz} and I.M.~{Khalatnikov}, \emph{{Investigations in relativistic cosmology{\textdagger}}}, \href{https://doi.org/10.1080/00018736300101283}{\emph{Advances in Physics} {\bfseries 12} (1963) 185}.

\bibitem{Weinberg:1972gravitation}
S.~Weinberg, \emph{Gravitation and Cosmology: Principles and Applications of the General Theory of Relativity}, Wiley (1972).

\bibitem{Peebles:1980LSS}
P.~Peebles, \emph{The Large-scale Structure of the Universe}, Princeton Series in Physics, Princeton University Press (1980).

\bibitem{MaBertschinger:1995}
C.-P.~{Ma} and E.~{Bertschinger}, \emph{{Cosmological Perturbation Theory in the Synchronous and Conformal Newtonian Gauges}}, \href{https://doi.org/10.1086/176550}{\emph{\apj} {\bfseries 455} (1995) 7} [\href{https://arxiv.org/abs/astro-ph/9506072}{{\ttfamily astro-ph/9506072}}].

\bibitem{Bernardeau:2002review}
F.~{Bernardeau}, S.~{Colombi}, E.~{Gazta{\~n}aga} and R.~{Scoccimarro}, \emph{{Large-scale structure of the Universe and cosmological perturbation theory}}, \href{https://doi.org/10.1016/S0370-1573(02)00135-7}{\emph{\physrep} {\bfseries 367} (2002) 1} [\href{https://arxiv.org/abs/astro-ph/0112551}{{\ttfamily astro-ph/0112551}}].

\bibitem{AnguloHahn:2022review}
R.E.~{Angulo} and O.~{Hahn}, \emph{{Large-scale dark matter simulations}}, \href{https://doi.org/10.1007/s41115-021-00013-z}{\emph{Living Reviews in Computational Astrophysics} {\bfseries 8} (2022) 1} [\href{https://arxiv.org/abs/2112.05165}{{\ttfamily 2112.05165}}].

\bibitem{CMBfast}
U.~{Seljak} and M.~{Zaldarriaga}, \emph{{A Line-of-Sight Integration Approach to Cosmic Microwave Background Anisotropies}}, \href{https://doi.org/10.1086/177793}{\emph{\apj} {\bfseries 469} (1996) 437} [\href{https://arxiv.org/abs/astro-ph/9603033}{{\ttfamily astro-ph/9603033}}].

\bibitem{WMAPmission:2003}
C.L.~{Bennett}, M.~{Bay}, M.~{Halpern}, G.~{Hinshaw}, C.~{Jackson}, N.~{Jarosik} et~al., \emph{{The Microwave Anisotropy Probe Mission}}, \href{https://doi.org/10.1086/345346}{\emph{\apj} {\bfseries 583} (2003) 1} [\href{https://arxiv.org/abs/astro-ph/0301158}{{\ttfamily astro-ph/0301158}}].

\bibitem{Planckmission:2011}
{Planck Collaboration}, P.A.R.~{Ade}, N.~{Aghanim}, M.~{Arnaud}, M.~{Ashdown}, J.~{Aumont} et~al., \emph{{Planck early results. I. The Planck mission}}, \href{https://doi.org/10.1051/0004-6361/201116464}{\emph{\aap} {\bfseries 536} (2011) A1} [\href{https://arxiv.org/abs/1101.2022}{{\ttfamily 1101.2022}}].

\bibitem{CAMB}
A.~{Lewis}, A.~{Challinor} and A.~{Lasenby}, \emph{{Efficient Computation of Cosmic Microwave Background Anisotropies in Closed Friedmann-Robertson-Walker Models}}, \href{https://doi.org/10.1086/309179}{\emph{\apj} {\bfseries 538} (2000) 473} [\href{https://arxiv.org/abs/astro-ph/9911177}{{\ttfamily astro-ph/9911177}}].

\bibitem{CLASS}
J.~{Lesgourgues}, \emph{{The Cosmic Linear Anisotropy Solving System (CLASS) I: Overview}}, \href{https://doi.org/10.48550/arXiv.1104.2932}{\emph{arXiv e-prints} (2011) arXiv:1104.2932} [\href{https://arxiv.org/abs/1104.2932}{{\ttfamily 1104.2932}}].

\bibitem{CLASS_approx}
D.~{Blas}, J.~{Lesgourgues} and T.~{Tram}, \emph{{The Cosmic Linear Anisotropy Solving System (CLASS). Part II: Approximation schemes}}, \href{https://doi.org/10.1088/1475-7516/2011/07/034}{\emph{\jcap} {\bfseries 2011} (2011) 034} [\href{https://arxiv.org/abs/1104.2933}{{\ttfamily 1104.2933}}].

\bibitem{Refregier:2018}
A.~{Refregier}, L.~{Gamper}, A.~{Amara} and L.~{Heisenberg}, \emph{{PyCosmo: An integrated cosmological Boltzmann solver}}, \href{https://doi.org/10.1016/j.ascom.2018.08.001}{\emph{Astronomy and Computing} {\bfseries 25} (2018) 38} [\href{https://arxiv.org/abs/1708.05177}{{\ttfamily 1708.05177}}].

\bibitem{Bucher:2000}
M.~{Bucher}, K.~{Moodley} and N.~{Turok}, \emph{{General primordial cosmic perturbation}}, \href{https://doi.org/10.1103/PhysRevD.62.083508}{\emph{\prd} {\bfseries 62} (2000) 083508} [\href{https://arxiv.org/abs/astro-ph/9904231}{{\ttfamily astro-ph/9904231}}].

\bibitem{EFTgravityCAMB:2014}
B.~{Hu}, M.~{Raveri}, N.~{Frusciante} and A.~{Silvestri}, \emph{{Effective field theory of cosmic acceleration: An implementation in CAMB}}, \href{https://doi.org/10.1103/PhysRevD.89.103530}{\emph{\prd} {\bfseries 89} (2014) 103530} [\href{https://arxiv.org/abs/1312.5742}{{\ttfamily 1312.5742}}].

\bibitem{HiClass:2017}
M.~{Zumalac{\'a}rregui}, E.~{Bellini}, I.~{Sawicki}, J.~{Lesgourgues} and P.G.~{Ferreira}, \emph{{hi\_class: Horndeski in the Cosmic Linear Anisotropy Solving System}}, \href{https://doi.org/10.1088/1475-7516/2017/08/019}{\emph{\jcap} {\bfseries 2017} (2017) 019} [\href{https://arxiv.org/abs/1605.06102}{{\ttfamily 1605.06102}}].

\bibitem{MGCAMB:2019}
A.~{Zucca}, L.~{Pogosian}, A.~{Silvestri} and G.B.~{Zhao}, \emph{{MGCAMB with massive neutrinos and dynamical dark energy}}, \href{https://doi.org/10.1088/1475-7516/2019/05/001}{\emph{\jcap} {\bfseries 2019} (2019) 001} [\href{https://arxiv.org/abs/1901.05956}{{\ttfamily 1901.05956}}].

\bibitem{AxionCAMB:2015}
R.~{Hlozek}, D.~{Grin}, D.J.E.~{Marsh} and P.G.~{Ferreira}, \emph{{A search for ultralight axions using precision cosmological data}}, \href{https://doi.org/10.1103/PhysRevD.91.103512}{\emph{\prd} {\bfseries 91} (2015) 103512} [\href{https://arxiv.org/abs/1410.2896}{{\ttfamily 1410.2896}}].

\bibitem{Foidl:2022}
H.~{Foidl} and T.~{Rindler-Daller}, \emph{{Cosmological structure formation in complex scalar field dark matter versus real ultralight axions: A comparative study using CLASS}}, \href{https://doi.org/10.1103/PhysRevD.105.123534}{\emph{\prd} {\bfseries 105} (2022) 123534} [\href{https://arxiv.org/abs/2203.09396}{{\ttfamily 2203.09396}}].

\bibitem{Hill:2020}
J.C.~{Hill}, E.~{McDonough}, M.W.~{Toomey} and S.~{Alexander}, \emph{{Early dark energy does not restore cosmological concordance}}, \href{https://doi.org/10.1103/PhysRevD.102.043507}{\emph{\prd} {\bfseries 102} (2020) 043507} [\href{https://arxiv.org/abs/2003.07355}{{\ttfamily 2003.07355}}].

\bibitem{JAX}
{Google Inc.}, ``Jax: High-performance array computing.'' \url{https://jax.readthedocs.io} Last accessed on 2023-07-20.

\bibitem{Kidger:2021}
P.~Kidger, \emph{On neural differential equations}, Ph.D. thesis, University of Oxford, 2021.

\bibitem{Pettinari:2013SONG}
G.W.~{Pettinari}, C.~{Fidler}, R.~{Crittenden}, K.~{Koyama} and D.~{Wands}, \emph{{The intrinsic bispectrum of the cosmic microwave background}}, \href{https://doi.org/10.1088/1475-7516/2013/04/003}{\emph{\jcap} {\bfseries 2013} (2013) 003} [\href{https://arxiv.org/abs/1302.0832}{{\ttfamily 1302.0832}}].

\bibitem{Wengert:1964}
R.E.~Wengert, \emph{A simple automatic derivative evaluation program}, \href{https://doi.org/10.1145/355586.364791}{\emph{Commun. ACM} {\bfseries 7} (1964) 463–464}.

\bibitem{Mathematica}
W.R.~Inc., ``Mathematica, {V}ersion 13.3.''

\bibitem{Sympy}
A.~Meurer, C.P.~Smith, M.~Paprocki, O.~\v{C}ert\'{i}k, S.B.~Kirpichev, M.~Rocklin et~al., \emph{Sympy: symbolic computing in python}, \href{https://doi.org/10.7717/peerj-cs.103}{\emph{PeerJ Computer Science} {\bfseries 3} (2017) e103}.

\bibitem{onken2020discretize}
D.~Onken and L.~Ruthotto, \emph{Discretize-optimize vs. optimize-discretize for time-series regression and continuous normalizing flows}, {\emph{Preprint} (2020) } [\href{https://arxiv.org/abs/2005.13420}{{\ttfamily 2005.13420}}].

\bibitem{flax2020github}
J.~Heek, A.~Levskaya, A.~Oliver, M.~Ritter, B.~Rondepierre, A.~Steiner et~al., \emph{{F}lax: A neural network library and ecosystem for {JAX}},  2023.

\bibitem{kaymak2023end}
M.C.~Kaymak, S.S.~Schoenholz, E.D.~Cubuk, K.A.~O’Hearn, K.M.~Merz~Jr and H.M.~Aktulga, \emph{End-to-end differentiable reactive molecular dynamics simulations using jax},  in \emph{International Conference on High Performance Computing}, pp.~202--219, Springer, 2023.

\bibitem{phan2019composable}
D.~Phan, N.~Pradhan and M.~Jankowiak, \emph{Composable effects for flexible and accelerated probabilistic programming in numpyro}, {\emph{arXiv preprint arXiv:1912.11554} (2019) } [\href{https://arxiv.org/abs/1912.11554}{{\ttfamily 1912.11554}}].

\bibitem{bingham2019pyro}
E.~Bingham, J.P.~Chen, M.~Jankowiak, F.~Obermeyer, N.~Pradhan, T.~Karaletsos et~al., \emph{Pyro: Deep universal probabilistic programming}, {\emph{J. Mach. Learn. Res.} {\bfseries 20} (2019) 28:1}.

\bibitem{campagne2023jax}
J.-E.~{Campagne}, F.~{Lanusse}, J.~{Zuntz}, A.~{Boucaud}, S.~{Casas}, M.~{Karamanis} et~al., \emph{{JAX-COSMO: An End-to-End Differentiable and GPU Accelerated Cosmology Library}}, \href{https://doi.org/10.21105/astro.2302.05163}{\emph{The Open Journal of Astrophysics} {\bfseries 6} (2023) 15} [\href{https://arxiv.org/abs/2302.05163}{{\ttfamily 2302.05163}}].

\bibitem{Piras23}
D.~{Piras} and A.~{Spurio Mancini}, \emph{{CosmoPower-JAX: high-dimensional Bayesian inference with differentiable cosmological emulators}}, \href{https://doi.org/10.21105/astro.2305.06347}{\emph{The Open Journal of Astrophysics} {\bfseries 6} (2023) 20} [\href{https://arxiv.org/abs/2305.06347}{{\ttfamily 2305.06347}}].

\bibitem{pmwd}
Y.~{Li}, L.~{Lu}, C.~{Modi}, D.~{Jamieson}, Y.~{Zhang}, Y.~{Feng} et~al., \emph{{pmwd: A Differentiable Cosmological Particle-Mesh $N$-body Library}}, \href{https://doi.org/10.48550/arXiv.2211.09958}{\emph{Preprint} (2022) arXiv:2211.09958} [\href{https://arxiv.org/abs/2211.09958}{{\ttfamily 2211.09958}}].

\bibitem{Li:2023ehn}
X.~Li, R.~Mandelbaum, M.~Jarvis, Y.~Li, A.~Park and T.~Zhang, \emph{{A Differentiable Perturbation-based Weak Lensing Shear Estimator}}, {\emph{Preprint} (2023) } [\href{https://arxiv.org/abs/2309.06506}{{\ttfamily 2309.06506}}].

\bibitem{Stevanovich:2023}
D.~Stevanovich, A.P.~Hearin and D.~Nagai, \emph{A differentiable model of the evolution of dark matter halo concentration}, {\emph{Preprint} (2023) } [\href{https://arxiv.org/abs/2309.07854}{{\ttfamily 2309.07854}}].

\bibitem{hairer2010solving}
E.~Hairer and G.~Wanner, \emph{Solving Ordinary Differential Equations II: Stiff and Differential-Algebraic Problems}, Springer Series in Computational Mathematics, Springer Berlin Heidelberg (2010).

\bibitem{Nadkarni-Ghosh:2017}
S.~{Nadkarni-Ghosh} and A.~{Refregier}, \emph{{The Einstein-Boltzmann equations revisited}}, \href{https://doi.org/10.1093/mnras/stx1662}{\emph{\mnras} {\bfseries 471} (2017) 2391} [\href{https://arxiv.org/abs/1612.06697}{{\ttfamily 1612.06697}}].

\bibitem{CLASS_ncdm}
J.~{Lesgourgues} and T.~{Tram}, \emph{{The Cosmic Linear Anisotropy Solving System (CLASS) IV: efficient implementation of non-cold relics}}, \href{https://doi.org/10.1088/1475-7516/2011/09/032}{\emph{\jcap} {\bfseries 2011} (2011) 032} [\href{https://arxiv.org/abs/1104.2935}{{\ttfamily 1104.2935}}].

\bibitem{kvaerno2004singly}
A.~Kv{\ae}rn{\o}, \emph{Singly diagonally implicit runge--kutta methods with an explicit first stage}, {\emph{BIT Numerical Mathematics} {\bfseries 44} (2004) 489}.

\bibitem{Howlett:2012}
C.~{Howlett}, A.~{Lewis}, A.~{Hall} and A.~{Challinor}, \emph{{CMB power spectrum parameter degeneracies in the era of precision cosmology}}, \href{https://doi.org/10.1088/1475-7516/2012/04/027}{\emph{\jcap} {\bfseries 2012} (2012) 027} [\href{https://arxiv.org/abs/1201.3654}{{\ttfamily 1201.3654}}].

\bibitem{Seager:1999recfast}
S.~{Seager}, D.D.~{Sasselov} and D.~{Scott}, \emph{{A New Calculation of the Recombination Epoch}}, \href{https://doi.org/10.1086/312250}{\emph{\apjl} {\bfseries 523} (1999) L1} [\href{https://arxiv.org/abs/astro-ph/9909275}{{\ttfamily astro-ph/9909275}}].

\bibitem{Peebles:1968recomb}
P.J.E.~{Peebles}, \emph{{Recombination of the Primeval Plasma}}, \href{https://doi.org/10.1086/149628}{\emph{\apj} {\bfseries 153} (1968) 1}.

\bibitem{Chluba:2011cosmorec}
J.~{Chluba} and R.M.~{Thomas}, \emph{{Towards a complete treatment of the cosmological recombination problem}}, \href{https://doi.org/10.1111/j.1365-2966.2010.17940.x}{\emph{\mnras} {\bfseries 412} (2011) 748} [\href{https://arxiv.org/abs/1010.3631}{{\ttfamily 1010.3631}}].

\bibitem{Ali-Haimoud:2011hyrec}
Y.~{Ali-Ha{\"\i}moud} and C.M.~{Hirata}, \emph{{HyRec: A fast and highly accurate primordial hydrogen and helium recombination code}}, \href{https://doi.org/10.1103/PhysRevD.83.043513}{\emph{\prd} {\bfseries 83} (2011) 043513} [\href{https://arxiv.org/abs/1011.3758}{{\ttfamily 1011.3758}}].

\bibitem{Lee:2020hyrec2}
N.~{Lee} and Y.~{Ali-Ha{\"\i}moud}, \emph{{HYREC-2: A highly accurate sub-millisecond recombination code}}, \href{https://doi.org/10.1103/PhysRevD.102.083517}{\emph{\prd} {\bfseries 102} (2020) 083517} [\href{https://arxiv.org/abs/2007.14114}{{\ttfamily 2007.14114}}].

\bibitem{Planck2018_cosmoparam}
{Planck Collaboration}, N.~{Aghanim}, Y.~{Akrami}, M.~{Ashdown}, J.~{Aumont}, C.~{Baccigalupi} et~al., \emph{{Planck 2018 results. VI. Cosmological parameters}}, \href{https://doi.org/10.1051/0004-6361/201833910}{\emph{\aap} {\bfseries 641} (2020) A6} [\href{https://arxiv.org/abs/1807.06209}{{\ttfamily 1807.06209}}].

\bibitem{Desjacques:2018biasreview}
V.~{Desjacques}, D.~{Jeong} and F.~{Schmidt}, \emph{{Large-scale galaxy bias}}, \href{https://doi.org/10.1016/j.physrep.2017.12.002}{\emph{\physrep} {\bfseries 733} (2018) 1} [\href{https://arxiv.org/abs/1611.09787}{{\ttfamily 1611.09787}}].

\bibitem{Kaiser:1987}
N.~{Kaiser}, \emph{{Clustering in real space and in redshift space}}, \href{https://doi.org/10.1093/mnras/227.1.1}{\emph{\mnras} {\bfseries 227} (1987) 1}.

\bibitem{Cole:1994rsdmultipole}
S.~{Cole}, K.B.~{Fisher} and D.H.~{Weinberg}, \emph{{Fourier Analysis of Redshift Space Distortions and the Determination of Omega}}, \href{https://doi.org/10.1093/mnras/267.3.785}{\emph{\mnras} {\bfseries 267} (1994) 785} [\href{https://arxiv.org/abs/astro-ph/9308003}{{\ttfamily astro-ph/9308003}}].

\bibitem{Hamilton:2000}
A.J.S.~{Hamilton}, \emph{{Uncorrelated modes of the non-linear power spectrum}}, \href{https://doi.org/10.1046/j.1365-8711.2000.03071.x}{\emph{\mnras} {\bfseries 312} (2000) 257} [\href{https://arxiv.org/abs/astro-ph/9905191}{{\ttfamily astro-ph/9905191}}].

\bibitem{Talman:1978}
J.D.~{Talman}, \emph{{Numerical Fourier and Bessel Transforms in Logarithmic Variables}}, \href{https://doi.org/10.1016/0021-9991(78)90107-9}{\emph{Journal of Computational Physics} {\bfseries 29} (1978) 35}.

\bibitem{AP:1979}
C.~{Alcock} and B.~{Paczynski}, \emph{{An evolution free test for non-zero cosmological constant}}, \href{https://doi.org/10.1038/281358a0}{\emph{\nat} {\bfseries 281} (1979) 358}.

\bibitem{Percival:2004}
W.J.~{Percival}, D.~{Burkey}, A.~{Heavens}, A.~{Taylor}, S.~{Cole}, J.A.~{Peacock} et~al., \emph{{The 2dF Galaxy Redshift Survey: spherical harmonics analysis of fluctuations in the final catalogue}}, \href{https://doi.org/10.1111/j.1365-2966.2004.08146.x}{\emph{\mnras} {\bfseries 353} (2004) 1201} [\href{https://arxiv.org/abs/astro-ph/0406513}{{\ttfamily astro-ph/0406513}}].

\bibitem{EuclidForecast}
{Euclid Collaboration}, A.~{Blanchard}, S.~{Camera}, C.~{Carbone}, V.F.~{Cardone}, S.~{Casas} et~al., \emph{{Euclid preparation. VII. Forecast validation for Euclid cosmological probes}}, \href{https://doi.org/10.1051/0004-6361/202038071}{\emph{\aap} {\bfseries 642} (2020) A191} [\href{https://arxiv.org/abs/1910.09273}{{\ttfamily 1910.09273}}].

\bibitem{Casas:2024}
S.~{Casas}, J.~{Lesgourgues}, N.~{Sch{\"o}neberg}, V.M.~{Sabarish}, L.~{Rathmann}, M.~{Doerenkamp} et~al., \emph{{Euclid: Validation of the MontePython forecasting tools}}, \href{https://doi.org/10.1051/0004-6361/202346772}{\emph{\aap} {\bfseries 682} (2024) A90} [\href{https://arxiv.org/abs/2303.09451}{{\ttfamily 2303.09451}}].

\bibitem{Tegmark1997}
M.~{Tegmark}, \emph{{Measuring Cosmological Parameters with Galaxy Surveys}}, \href{https://doi.org/10.1103/PhysRevLett.79.3806}{\emph{\prl} {\bfseries 79} (1997) 3806} [\href{https://arxiv.org/abs/astro-ph/9706198}{{\ttfamily astro-ph/9706198}}].

\bibitem{Amendola2013}
L.~{Amendola}, S.~{Appleby}, D.~{Bacon}, T.~{Baker}, M.~{Baldi}, N.~{Bartolo} et~al., \emph{{Cosmology and Fundamental Physics with the Euclid Satellite}}, \href{https://doi.org/10.12942/lrr-2013-6}{\emph{Living Reviews in Relativity} {\bfseries 16} (2013) 6} [\href{https://arxiv.org/abs/1206.1225}{{\ttfamily 1206.1225}}].

\bibitem{Bhandari2021}
N.~{Bhandari}, C.D.~{Leonard}, M.M.~{Rau} and R.~{Mandelbaum}, \emph{{Fisher Matrix Stability}}, \href{https://doi.org/10.48550/arXiv.2101.00298}{\emph{arXiv e-prints} (2021) arXiv:2101.00298} [\href{https://arxiv.org/abs/2101.00298}{{\ttfamily 2101.00298}}].

\bibitem{PeacockDodds:1994}
J.A.~{Peacock} and S.J.~{Dodds}, \emph{{Reconstructing the Linear Power Spectrum of Cosmological Mass Fluctuations}}, \href{https://doi.org/10.1093/mnras/267.4.1020}{\emph{\mnras} {\bfseries 267} (1994) 1020} [\href{https://arxiv.org/abs/astro-ph/9311057}{{\ttfamily astro-ph/9311057}}].

\bibitem{ChavesMontero:2018}
J.~{Chaves-Montero}, R.E.~{Angulo} and C.~{Hern{\'a}ndez-Monteagudo}, \emph{{The effect of photometric redshift uncertainties on galaxy clustering and baryonic acoustic oscillations}}, \href{https://doi.org/10.1093/mnras/sty924}{\emph{\mnras} {\bfseries 477} (2018) 3892} [\href{https://arxiv.org/abs/1610.09688}{{\ttfamily 1610.09688}}].

\bibitem{Archidiacono:2024}
{Euclid Collaboration}, M.~{Archidiacono}, J.~{Lesgourgues}, S.~{Casas}, S.~{Pamuk}, N.~{Sch{\"o}neberg} et~al., \emph{{Euclid preparation. Sensitivity to neutrino parameters}}, \href{https://doi.org/10.48550/arXiv.2405.06047}{\emph{arXiv e-prints} (2024) arXiv:2405.06047} [\href{https://arxiv.org/abs/2405.06047}{{\ttfamily 2405.06047}}].

\bibitem{Seo2003}
H.-J.~{Seo} and D.J.~{Eisenstein}, \emph{{Probing Dark Energy with Baryonic Acoustic Oscillations from Future Large Galaxy Redshift Surveys}}, \href{https://doi.org/10.1086/379122}{\emph{\apj} {\bfseries 598} (2003) 720} [\href{https://arxiv.org/abs/astro-ph/0307460}{{\ttfamily astro-ph/0307460}}].

\bibitem{Ivanov:2022}
M.M.~{Ivanov}, \emph{{Effective Field Theory for Large Scale Structure}}, \href{https://doi.org/10.48550/arXiv.2212.08488}{\emph{arXiv e-prints} (2022) arXiv:2212.08488} [\href{https://arxiv.org/abs/2212.08488}{{\ttfamily 2212.08488}}].

\bibitem{Rampf:2021sc}
C.~{Rampf} and O.~{Hahn}, \emph{{Shell-crossing in a {\ensuremath{\Lambda}}CDM Universe}}, \href{https://doi.org/10.1093/mnrasl/slaa198}{\emph{\mnras} {\bfseries 501} (2021) L71} [\href{https://arxiv.org/abs/2010.12584}{{\ttfamily 2010.12584}}].

\bibitem{Schmidt:2021}
F.~{Schmidt}, \emph{{An n-th order Lagrangian forward model for large-scale structure}}, \href{https://doi.org/10.1088/1475-7516/2021/04/033}{\emph{\jcap} {\bfseries 2021} (2021) 033} [\href{https://arxiv.org/abs/2012.09837}{{\ttfamily 2012.09837}}].

\bibitem{Arico:2021baccoemu}
G.~{Aric{\`o}}, R.E.~{Angulo} and M.~{Zennaro}, \emph{{Accelerating Large-Scale-Structure data analyses by emulating Boltzmann solvers and Lagrangian Perturbation Theory}}, \href{https://doi.org/10.48550/arXiv.2104.14568}{\emph{arXiv e-prints} (2021) arXiv:2104.14568} [\href{https://arxiv.org/abs/2104.14568}{{\ttfamily 2104.14568}}].

\bibitem{Bonici:2022emu}
M.~{Bonici}, L.~{Biggio}, C.~{Carbone} and L.~{Guzzo}, \emph{{Fast emulation of two-point angular statistics for photometric galaxy surveys}}, \href{https://doi.org/10.48550/arXiv.2206.14208}{\emph{arXiv e-prints} (2022) arXiv:2206.14208} [\href{https://arxiv.org/abs/2206.14208}{{\ttfamily 2206.14208}}].

\bibitem{Spurio:2022}
A.~{Spurio Mancini}, D.~{Piras}, J.~{Alsing}, B.~{Joachimi} and M.P.~{Hobson}, \emph{{COSMOPOWER: emulating cosmological power spectra for accelerated Bayesian inference from next-generation surveys}}, \href{https://doi.org/10.1093/mnras/stac064}{\emph{\mnras} {\bfseries 511} (2022) 1771} [\href{https://arxiv.org/abs/2106.03846}{{\ttfamily 2106.03846}}].

\bibitem{Donald-Mccann:2022}
J.~{Donald-McCann}, F.~{Beutler}, K.~{Koyama} and M.~{Karamanis}, \emph{{MATRYOSHKA: halo model emulator for the galaxy power spectrum}}, \href{https://doi.org/10.1093/mnras/stac239}{\emph{\mnras} {\bfseries 511} (2022) 3768} [\href{https://arxiv.org/abs/2109.15236}{{\ttfamily 2109.15236}}].

\bibitem{Bonici:2023diffemu}
M.~{Bonici}, F.~{Bianchini} and J.~{Ruiz-Zapatero}, \emph{{Capse.jl: efficient and auto-differentiable CMB power spectra emulation}}, \href{https://doi.org/10.48550/arXiv.2307.14339}{\emph{arXiv e-prints} (2023) arXiv:2307.14339} [\href{https://arxiv.org/abs/2307.14339}{{\ttfamily 2307.14339}}].

\bibitem{Chisari:2019:CCL}
N.E.~{Chisari}, D.~{Alonso}, E.~{Krause}, C.D.~{Leonard}, P.~{Bull}, J.~{Neveu} et~al., \emph{{Core Cosmology Library: Precision Cosmological Predictions for LSST}}, \href{https://doi.org/10.3847/1538-4365/ab1658}{\emph{\apjs} {\bfseries 242} (2019) 2} [\href{https://arxiv.org/abs/1812.05995}{{\ttfamily 1812.05995}}].

\bibitem{Ballesteros:2010}
G.~{Ballesteros} and J.~{Lesgourgues}, \emph{{Dark energy with non-adiabatic sound speed: initial conditions and detectability}}, \href{https://doi.org/10.1088/1475-7516/2010/10/014}{\emph{\jcap} {\bfseries 2010} (2010) 014} [\href{https://arxiv.org/abs/1004.5509}{{\ttfamily 1004.5509}}].

\bibitem{Chevallier:2001}
M.~{Chevallier} and D.~{Polarski}, \emph{{Accelerating Universes with Scaling Dark Matter}}, \href{https://doi.org/10.1142/S0218271801000822}{\emph{International Journal of Modern Physics D} {\bfseries 10} (2001) 213} [\href{https://arxiv.org/abs/gr-qc/0009008}{{\ttfamily gr-qc/0009008}}].

\bibitem{Linder:2003}
E.V.~{Linder}, \emph{{Exploring the Expansion History of the Universe}}, \href{https://doi.org/10.1103/PhysRevLett.90.091301}{\emph{\prl} {\bfseries 90} (2003) 091301} [\href{https://arxiv.org/abs/astro-ph/0208512}{{\ttfamily astro-ph/0208512}}].

\end{thebibliography}\endgroup

\appendix

\section{Governing equations}
%%%%%%%%%%%%%%%%%%%%%%%%%%%%%%%%%%%%%%%%%%%%%%%%%%%%%%%%%%%%%%%%%
\label{sec:governing_equations}
In the {\codename} Einstein-Boltzmann solver, we currently implement the linearised equations for perturbations of the metric, as we all as CDM, baryons, relativistic species, massive neutrinos, and a dark energy fluid, which we briefly summarise here. Specifically we have implemented the following set of equations, in the {\em synchronous} gauge, that we reproduce here for reference. We largely follow \cite{MaBertschinger:1995} and use their notation, except that we denote derivatives w.r.t. conformal time by a dash rather than a dot. 

\paragraph{Metric perturbations. } In the synchronous gauge, coordinates are defined by freely falling observers, and scalar metric perturbations are described by two scalar potentials $\eta$ and $h$, cf. \cite{MaBertschinger:1995}. Their evolution is governed by the Einstein equations, which in Fourier space read
\begin{subequations}
\begin{align}
    k^2\eta - \frac{1}{2}\mathcal{H}h' &= 4\pi G a^2 \;\delta T^0_{\phantom{0}0}  \\
    k^2\eta' &= 4\pi G a^2 (\overline{\rho}+\overline{P})\;\theta \\
    h'' + 2\mathcal{H}h' - 2k^2\eta &= -8\pi G a^2 \delta T^i_{\phantom{i}i} \\
    h'' + 6 \eta'' + 2\mathcal{H}\left(h'+6\eta'\right) - 2k^2\eta &= -24\pi G a^2 (\overline{\rho}+\overline{P})\;\sigma.
  \end{align}
\end{subequations}
These equations are overdetermined so that e.g. $h'$ is not a dynamical degree of freedom.

\paragraph{CDM.}
As coordinates are defined by freely falling observers, the equations of motion of collisionless cold dark matter are particularly simple:
\begin{subequations}
    \begin{align}
        \delta_c' &= - \frac{h'}{2} \\
        \theta_c &= \sigma_c = 0\;.
      \end{align}
\end{subequations}

\paragraph{Baryons.} In contrast, baryons have a finite sound speed and couple to the photons via Thomson scattering while free electrons are present. The temperature and ionization state are determined at the background level by the thermal history solver. The baryon equations of motion (EOMs) are then given by 
\begin{subequations}
    \begin{align}
        \delta_b' &= -\theta_b - \frac{h'}{2} \\
        \theta_b' &= -\mathcal{H}\theta_b + c_s^2k^2\delta_b + \frac{4}{3} \frac{\overline{\rho}_\gamma}{\overline{\rho}_b}a n_e \sigma_T (\theta_\gamma-\theta_b)\\
        \intertext{where $n_e$ is the homogeneous mean electron density, $\sigma_T$ is the Thomson scattering cross section, and $c_s$ the adiabatic sound speed of the baryon fluid. The latter is given by}
        c_s^2 &= \frac{k_B T_b}{\mu}\left(1-\frac{1}{3}\frac{\dd \log T_b}{\dd \log a}\right)\\
        \intertext{where}
        T_b' &= -2\mathcal{H}T_b + \frac{8}{3}\frac{\mu}{m_e}\frac{\overline{\rho}_\gamma}{\overline{\rho}_b} a n_e \sigma_T (T_\gamma-T_b)
      \end{align}
      is the evolution of the baryon temperature, which explicitly depends on the mean molecular weight $\mu$ and electron density, which are computed by the recombination solver. The ratio $\overline{\rho}_\gamma/\overline{\rho}_b$ denotes the photon-to-baryon ratio.
\end{subequations}

\paragraph{Photons.} Photons follow Bose-Einstein statistics with two polarisation states, and are coupled to the baryons via Thomson scattering off free electrons. Their EOMs are given by a Boltzmann hierarchy of moments. The first three moments (corresponding to density, velocity, and shear perturbations) are given by
\begin{subequations}
    \begin{align}
        \delta_\gamma' &= -\frac{4}{3}\theta_\gamma - \frac{2}{3}h' \\
        \theta_\gamma' &= k^2\left( \frac{1}{4}\delta_\gamma - \sigma_\gamma \right) + a n_e \sigma_T\left(\theta_b-\theta_\gamma\right)\\
       2 \sigma_\gamma' = F_{\gamma}^{(2)\prime} &= \frac{8}{15}\theta_\gamma - \frac{3}{5}k F_{\gamma}^{(3)} + \frac{4}{15} h' + \frac{8}{5} \eta' - \frac{9}{5}a n_e \sigma_T \sigma_\gamma + \frac{1}{10}a n_e \sigma_T(G_{\gamma}^{(0)}+G_{\gamma}^{(2)})\\
       \intertext{The higher Boltzmann moments for $\ell_\text{max}>\ell\ge3$ are given by}
        {F}_{\gamma}^{(\ell)\prime} &= \frac{k}{2\ell+1}\left[\ell F_{\gamma}^{(\ell-1)}-(\ell+1)F_{\gamma}^{(\ell+1)}\right] - a n_e \sigma_T F_{\gamma}^{(\ell)}\\
        \intertext{and the hierarchy is truncated by hand as in \cite{MaBertschinger:1995} by setting}
        {F}_{\gamma}^{(\ell_\text{max})\prime} &\approx k F_{\gamma}^{(\ell_\text{max}-1)} - \frac{\ell_\text{max}+1}{\tau}F_{\gamma}^{(\ell_\text{max})} - a n_e \sigma_T F_{\gamma}^{(\ell_\text{max})}. \\
        \intertext{The polarisation moments for $0\le\ell<\ell_\text{max}$ are given by}
        {G}_{\gamma}^{(\ell)\prime} &= \frac{k}{2\ell+1}\left[\ell G_{\gamma}^{(\ell-1)}-(\ell+1)G_{\gamma}^{(\ell+1)}\right] + \nonumber \\
        &\quad + a n_e \sigma_T \left[-G_{\gamma}^{(\ell)}+\frac{1}{2}\left(F_{\gamma}^{(2)}+G_{\gamma}^{(0)}+G_{\gamma}^{(2)}\right)\left(\delta_{\ell0}+\frac{1}{5}\delta_{\ell2}\right)\right]\;, \\
        \intertext{(where $\delta_{\ell j}$ is a Kronecker symbol). The respective hierarchy truncation is given by}
        {G}_{\gamma}^{(\ell_\text{max})\prime} &\approx k G_{\gamma}^{(\ell_\text{max}-1)} - \frac{\ell_\text{max}+1}{\tau}G_{\gamma}^{(\ell_\text{max})} - a n_e \sigma_T G_{\gamma}^{(\ell_\text{max})}\;.
      \end{align}
\end{subequations}

\paragraph{Massless neutrinos.} Massless neutrinos follow Fermi-Dirac statistics, with EOMs given by a Boltzmann hierarchy of moments, of which the first three are
\begin{subequations}
\begin{align}
  {\delta}_\nu' &= -\frac{4}{3}\theta_\nu - \frac{2}{3}h' \\
  {\theta}_\nu' &= k^2\left( \frac{1}{4}\delta_\nu - \sigma_\nu \right) \\ 
  2 {\sigma}_\nu' = {F}_{\nu}^{(2)\prime} &= \frac{8}{15}\theta_\nu - \frac{3}{5}k F_{\nu}^{(3)} + \frac{4}{15} h' + \frac{8}{5} \eta'\;.
  \intertext{The higher Boltzmann moments for $3\le \ell < \ell_\text{max}$ are given by}
  {F}_{\nu}^{(\ell)\prime} &= \frac{k}{2\ell+1}\left[\ell F_{\nu}^{(\ell-1)}-(\ell+1)F_{\nu}^{(\ell+1)}\right] \\
  \intertext{with a truncation given by}
  F_{\nu}^{(\ell_\text{max})\prime} &\approx kF_\nu^{(\ell_\text{max}-1)} -\frac{\ell_\text{max}+1}{\tau}F_\nu^{(\ell_\text{max})}\;.
  %% F_{\nu}^{(\ell_\text{max}+1)} &\approx \frac{(2\ell_\text{max}+1)}{k\tau} F_{\nu}^{(\ell_\text{max})} - F_{\nu}^{(\ell_\text{max}-1)}  
\end{align}
\end{subequations}

\paragraph{Massive neutrinos.}
Finally, for massive neutrinos, a momentum-dependent Boltzmann hierarchy has to be evolved. This is done by discretising the momentum space into $n$ bins with momenta $q\in \{q_0,\dots,q_n\}$, and evolving the hierarchy for each bin. The moments of the neutrino distribution function $\psi^{(\ell)}_q$ are then given for each momentum bin, i.e.
\begin{subequations}
\begin{align} 
  {\psi}^{(0)\prime}_q &= -\frac{qk}{\epsilon}\psi^{(1)}_q + \frac{1}{6}h'\frac{\dd \log f_0}{\dd \log q}\;, \\
  {\psi}^{(1)\prime}_q &= \frac{qk}{3\epsilon}\left(\psi^{(0)}_q - 2\psi^{(2)}_q\right)\;,\\
  {\psi}^{(2)\prime}_q &= \frac{qk}{5\epsilon}\left(2\psi^{(1)}_q - 3\psi^{(3)}_q\right) - \left(\frac{1}{15}h' + \frac{2}{5}\eta'\right)\frac{\dd \log f_0}{\dd \log q}\;,\\
  \intertext{where $\epsilon^2:=q^2+a^2\,m_\nu^2$. The higher Boltzmann moments for $3\le \ell < \ell_\text{max}$ are given by}
  {\psi}^{(\ell)\prime}_q &= \frac{qk}{(2\ell+1)\epsilon}\left(\ell\psi^{(\ell-1)}_q - (\ell+1)\psi^{(\ell+1)}_q\right)\;,\\
  \intertext{and the hierarchy is truncated as}
  \psi^{(\ell_\text{max})\prime}_q &\approx \frac{qk}{\epsilon} \psi^{(\ell_\text{max}-1)}_q - \frac{\ell_\text{max}+1}{\tau}\psi^{(\ell_\text{max})}_q\;.
  %%%\psi^{(\ell_\text{max}+1)}_q &\approx \frac{(2\ell_\text{max}+1)\epsilon}{qk\tau} \psi^{(\ell_\text{max})}_q - \psi^{(\ell_\text{max}-1)}\;.
\end{align}
The massive neutrino perturbations entering the field equations are then given in terms of the lowest moments as
\begin{align}
  \delta\rho_\nu &= \frac{4\pi}{a^4}\int \dd q\;q^2\epsilon f_0(q)\psi_0(q)\;,\\
  \delta P_\nu &= \frac{4\pi}{3a^4}\int \dd q\;\frac{q^4}{\epsilon} f_0(q)\psi_1(q)\;,\\
  (\overline{\rho}_\nu+\overline{P}_\nu)\theta_\nu &= \frac{4\pi k}{a^4}\int \dd q\;q^3 f_0(q)\psi_1(q)\;,\\
  (\overline{\rho}_\nu+\overline{P}_\nu)\sigma_\nu &= \frac{8\pi}{3a^4}\int \dd q\;\frac{q^4}{\epsilon} f_0(q)\psi_2(q)\;,
\end{align}
where $f_0(q)$ is the background neutrino distribution function, which is assumed to be Fermi-Dirac with zero chemical potential. The momentum bins are chosen in order to approximate well these integrals over the Fermi-Dirac distribution, cf. \cite{Howlett:2012} and \autoref{sec:approx}.
\end{subequations}

Finally, in the presence of massive neutrinos, the effective degrees of freedom of the ultrarelativistic species are modified. We follow \cite{Chisari:2019:CCL} (their eq.~84) and set the effective number of relativistic species given $N_\text{eff}$ as
\begin{align}
    N_{\nu,\text{rel}} = N_\text{eff} - T_{\nu,\text{massive}}^4 \left(\tfrac{4}{11}\right)^{-4/3}N_{\nu,\text{massive}},
\end{align}
where in our case $N_{\nu,\text{massive}}=1$ for one single species of massive neutrino currently. The massive neutrino temperature $T_{\nu,\text{massive}}$ is by default set to the temperature of the massless neutrinos $T_{\nu,\text{massive}} = \left(4/11\right)^{1/3}\,T_\text{CMB}$ so  that in this case $N_{\nu,\text{rel}}=N_\text{eff}-N_{\nu,\text{massive}}$. Note that $T_{\nu,\text{massive}}$ corresponds to \textsc{Class}' \texttt{T\_ncdm} parameter.

\paragraph{Quintessence/dark energy.} Finally, we allow for a dark energy perturbation, which is implemented following \cite{Ballesteros:2010}. We evolve the following perturbation variables
\begin{subequations}
    \begin{align}
        \delta_Q'&= -(1+w) (\theta_Q + \tfrac{1}{2}h') - 3 \mathcal{H} (c_a^2 - w) \delta_Q -9\mathcal{H}(1+w)(c_a^2-c_w^2)\frac{\theta_Q}{k^2}\\
        \theta_Q' &= -\mathcal{H} (1-3c_a^2) \theta_Q + \frac{c_a^2}{1+w} k^2 \delta_Q \;,
    \end{align}
\end{subequations}
where we assume the usual (Chevallier-Polarski-Linder \cite{Chevallier:2001,Linder:2003}) parameterisation of the equation of state parameter $w:=w_0+w_a(1-a)$, $c_a^2$ is the (constant) adiabatic sound speed of the dark energy perturbation, and $c_w^2 := w - w'/(3\mathcal{H}(1+w))$.

\section{Comparison with \textsc{Class} and \textsc{Camb} at early times }

\begin{figure}[h]
    \centering
    \includegraphics[width=0.99\textwidth]{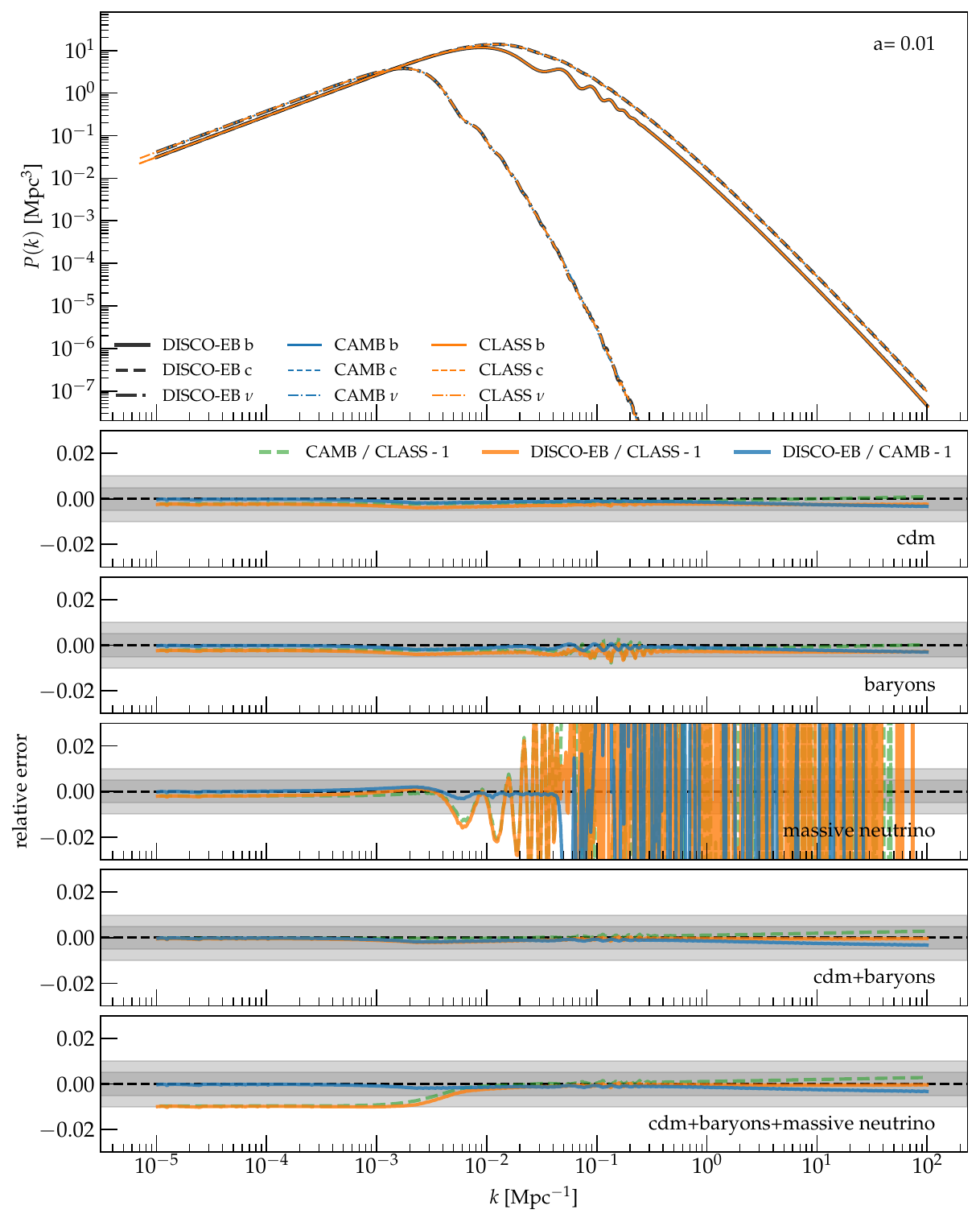}
    \caption{Same as \autoref{fig:comp_CLASS_z0}, but for $z = 99$ ($a = 0.01$).}
    \label{fig:comp_CLASS_z99}
\end{figure}

In \autoref{fig:comp_CLASS_z99}, we show the same comparison as in \autoref{fig:comp_CLASS_z0}, but for $z = 99$ ($a = 0.01$). The agreement is very good, with the exception of the massive neutrino perturbations where differences are large for wave numbers suppressed by free-streaming damping. Those scales are however dynamically unimportant. Minor differences are seen in the baryon perturbations, likely caused by our simplified thermal history solver. Finally, we see a difference between the \textsc{Camb} and \textsc{Class} cdm+baryon+massive neutrino spectrum for $k\lesssim 10^{-2} \,\text{Mpc}^{-1}$, which is arguably due to differences in how massive neutrinos are weighted in the total matter output spectrum in \textsc{Class} compared to \textsc{Camb} and \codenameEB{}.

\end{document}